\documentclass[fleqn,usenatbib]{mnras}
\usepackage{multirow}
\usepackage{newtxtext,newtxmath}

\usepackage[T1]{fontenc}
\DeclareRobustCommand{\VAN}[3]{#2}
\let\VANthebibliography\thebibliography
\def\thebibliography{\DeclareRobustCommand{\VAN}[3]{##3}\VANthebibliography}

\usepackage[pagewise]{lineno}

\usepackage{graphicx}	
\usepackage{amsmath}	
\usepackage{comment}

\title[Optical and X-ray studies of IGR J06074+2205]{Optical and X-ray Studies of the Be/X-ray Binary IGR J06074+2205}

\author[Chhotaray et al.]{
Birendra Chhotaray,$^{1,2}$\thanks{E-mail: birendra@prl.res.in, rsbirendra786@gmail.com}
Sachindra Naik$^{1}$,
Gaurava K. Jaisawal$^{3}$,
Goldy Ahuja$^{1,2}$
\\
$^{1}$Astronomy and Astrophysics Division, Physical Research Laboratory, Navrangpura, Ahmedabad - 380009, Gujarat, India\\
$^{2}$Indian Institute of Technology Gandhinagar, Palaj, Gandhinagar - 382055, Gujarat, India\\
$^{3}$DTU Space, Technical University of Denmark, Elektrovej 327-328, DK-2800 Lyngby, Denmark
}

\date{Accepted XXX. Received YYY; in original form ZZZ}

\pubyear{2024}

\begin{document}
\label{firstpage}
\pagerange{\pageref{firstpage}--\pageref{lastpage}}
\maketitle

\begin{abstract}
We present the results obtained from X-ray and optical analysis of the Be/X-ray binary IGR~J06074+2205, focusing on before, during, and after the X-ray outbursts in October and December 2023. The properties of the neutron star in the binary are investigated using NICER and NuSTAR observations during the X-ray outbursts. The pulse profiles across a broad energy range, are found to be strongly dependent on luminosity and energy, revealing the complex nature of the emitting region. An absorbed power-law can describe each NICER spectrum in the 1-7~keV band. The 3–79~keV NuSTAR spectrum can be well-described by a negative and positive power-law with an exponential cut-off model. Utilizing the MAXI/GSC long-term light curve, we estimate the probable orbital period to be 80 or 80/n (n=2,3,4) days. We investigate the evolution of the circumstellar disc around the Be star by using optical spectroscopic observations of the system between 2022 and 2024. We observe variable H$\alpha$ and FeII emission lines with an increase in equivalent width, indicating the presence of a dynamic circumstellar disc.  A distinct variation in the V/R value for H$\alpha$ and FeII lines is also observed. The appearance of additional emission lines, such as HeI (5875.72~\AA), HeI (6678~\AA), and HeI (7065~\AA), during the post-outburst observation in February 2024 suggests the growing of a larger or denser circumstellar disc. The disc continues to grow without any noticeable mass loss, even during the 2023 X-ray outbursts, which may lead to a future giant X-ray outburst.
\end{abstract}




\section{Introduction}

Be/X-ray binaries (BeXRBs) constitute the largest group within the high mass X-ray binaries (HMXBs). These binaries typically host a neutron star as the compact object, coupled with a massive non-supergiant star (with a mass range of 10-20 M$_{\odot}$) serving as the optical companion, orbiting around their common center of mass (e.g.,~ \citealt{2011BASI...39..429P, 2011Ap&SS.332....1R}). The neutron stars within these binary systems derive their energy by accreting mass from the circumstellar disc of the Be stars. Notably, the companion Be stars in BeXRBs are bright in optical and infrared wavebands. These stars typically belong to the early-B or late-O spectral types and frequently display distinct emission lines such as HI, HeI, and FeII at certain stages of their evolution~\citep{2003PASP..115.1153P}. The emission lines show variable shape and strength, which depend upon the physical properties, orientation, and interaction with the compact object of the circumstellar disc~\citep{2006MNRAS.368..447C,2013PASJ...65...83M,2023MNRAS.518.5089C,2024BSRSL..93..657N}.

BeXRBs exhibit two distinct categories of X-ray activities: normal (Type~I) and giant (Type~II) X-ray outbursts \citep{1986ApJ...308..669S,1998A&A...338..505N}. Normal outbursts manifest as periodic or quasi-periodic events, typically occurring near the periastron passage of the orbiting neutron star. These outbursts last for a fraction (approximately 20\%) of the orbital period, reaching peak luminosities up to 10$^{37}$ erg s$^{-1}$. In contrast, giant X-ray outbursts are rare and irregular, and their infrequent occurrences are devoid of any orbital modulation. They persist for multiple orbital periods, showing peak luminosities equal to or exceeding $L_{\rm X}$ $\geq$10$^{37}$ erg s$^{-1}$ (\citealt{2011BASI...39..429P,2011Ap&SS.332....1R} \& references therein). The BeXRBs exhibit both short-term (e.g., due to pulsation of neutron star) and long-term (e.g., due to mass accretion close to the periastron passage of the neutron star) X-ray variabilities. The continuum emission from the BeXRB pulsar generally falls in the 0.2-100 keV energy range~\citep{2007ApJ...654..435B}. Apart from the continuum, the energy spectrum of pulsars also exhibits emission features mostly from iron (Fe), and absorption features in the 10-100 keV range caused by cyclotron resonant scattering \citep{1978ApJ...219L.105T, 2019A&A...622A..61S}. Detection of the cyclotron resonant scattering features (CRSFs) in the pulsars is a direct and most accurate method for estimating the surface magnetic field of the neutron stars.

IGR~J06074+2205, identified as a BeXRB, was initially discovered by the JEM-X telescope onboard INTEGRAL in 2003~\citep{2004ATel..223....1C}.  The source exhibited a flux of approximately 7~mCrab in the 3-10~keV range and about 15~mCrab within the 10-20~keV energy band. Additionally, a potential association between the radio source NVSS~J060718+220452 and IGR~J06074+2205 was suggested by \citet{2004ATel..226....1P}. Subsequently, \citet{2005ATel..682....1H} reported on the potential optical counterpart of IGR~J06074+2205. Their observations suggested the optical counterpart to be a Be star, as indicated by the presence of the H$\alpha$ emission line with an equivalent width of -6.6 \AA. Subsequently,  IGR~J06074+2205 was identified as a Be/X-ray binary by confirming the association of Be star and the X-ray source using the high spatial resolution capability of Chandra X-ray observatory \citep{2006ATel..959....1T}. \citealt{2010A&A...522A.107R} conducted an extensive analysis of the optical counterpart in the 4000-7000~\AA~range between 2006 and 2008. They suggested the optical counterpart of IGR~J06074+2205 to be a B0.5Ve spectral type star with a V-band magnitude of 12.3, situated at a distance of 4.5 kpc. However, recent studies by \citet{2022A&A...665A..69F} suggested the source distance to be 5.99 kpc using {\it Gaia}. From the analysis of the H$\alpha$ line, \citet{2010A&A...522A.107R} found  V/R (the ratio of violet-side to red-side peak intensities above continuum in units of continuum intensity representing a measure of the asymmetry of the line) variability, indicating significant changes in the structure of the circumstellar disc on timescales of months. Additionally, a decline in the equivalent width of the H$\alpha$ line over approximately 3 years was noted. Furthermore, the H$\alpha$ line feature in absorption was observed in March 2010, suggesting a possible disc loss event. Following this disc loss episode, a new decretion disc (characterized by H$\alpha$ emission) was detected during the observations conducted between 14 and 24 April 2012, as reported by \citet{2012ATel.4172....1S}. \citet{2024ATel16560....1S} carried out photometric observations in the optical waveband from 3 December 2023 to 17 March 2024 and found no optical variability associated with the X-ray outburst on 15 March 2024. Using photometric studies, \citet{2024OEJV..249....1N} suggested the precession or propagation of density wave in the circumstellar disc gives rise to optical variability at a timescale of 620 days. 

\citet{2018A&A...613A..52R} for the first time detected coherent X-ray pulsations from the X-ray source using the XMM-Newton observation on 29 September 2017. The pulsation was observed at 373.2 s at a low luminosity level of $\approx$ 10$^{34}$ erg s$^{-1}$ for a distance of 4.5 kpc. Interestingly, \citet{2018A&A...613A..52R} detected the pulsations during the disc-loss phase of the Be star. Their spectral analysis in the 0.4-12 keV range was characterized by a combination of an absorbed power law and thermal (blackbody) components, with the following optimal parameters: N$_{H}$ = (6.2 $\pm$ 0.5) $\times$ 10$^{21}$ cm$^{-2}$, kT$_{bb}$ = 1.16 $\pm$ 0.03 keV, and $\Gamma$ = 1.5 $\pm$ 0.1. The absorbed X-ray luminosity was determined to be L$_{X}$ = 1.4 $\times$ 10$^{34}$ erg s$^{-1}$, assuming a distance of 4.5 kpc. Recently, \citet{2024arXiv240217382R} reported the detection of a cyclotron absorption line at $\approx$51 keV, which corresponds to a magnetic field strength of 4$\times$10$^{12}$ G.

\begin{figure*}
    \includegraphics[trim={0 0cm 0 0cm},scale=0.8]{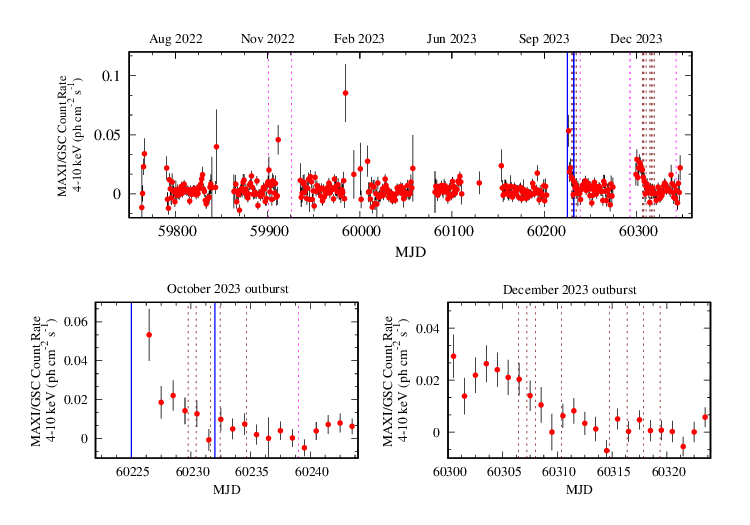}
    \caption{ Upper panel: One-day binned MAXI/GSC light curve of IGR~J06074+2205 in the 4-10 keV energy band between May 2022 and February 2024. Lower panels: MAXI/GSC count rate variations during the October and December 2023 X-ray outbursts. Vertical dashed lines indicate the epochs of NuSTAR (blue), NICER (brown), and optical (magenta) observations.}
    \label{fig:log}
\end{figure*}

In this work, we discuss the properties of IGR~J06074+2205 using X-ray and optical observations of the binary. As shown in Figure~\ref{fig:log}, we used X-ray data obtained during the X-ray outbursts in October and December 2023.  The optical observations were carried out between November 2022 and February 2024, covering both October and December 2023 X-ray outbursts, using the Mount Abu Infrared Observatory (MIRO) and the Indian Astronomical Observatory (IAO). The paper is structured as follows: Section~\ref{sec:2 X-ray} provides an overview of the observations and data reduction procedures applied to NuSTAR and NICER data. Section~\ref{sec:3 optical reduc} details the observations and data reduction procedures for MIRO and IAO data. Sections~\ref{sec:xray_timing} and \ref{sec:xray_spectral} delve into the results of X-ray timing and spectral analyses, respectively. Following this, Section~\ref{sec:optical analysis} presents findings from optical spectroscopic analysis. Discussion and conclusions are presented in Sections~\ref{discussion} and ~\ref{conclusion}, respectively. 

\section{X-ray observations \& data reduction}
\label{sec:2 X-ray}
\subsection{NuSTAR}
NuSTAR is a grazing incidence hard X-ray focusing telescope sensitive in 3–79 keV energy range \citep{2013ApJ...770..103H}. It consists of two identical focal plane modules: Focal Plane Module A (FPMA) and Focal Plane Module B (FPMB), where data from each module are processed separately. IGR~J06074+2204 was observed at  two epochs on MJD 60225 (ID 90901331002) and MJD 60232 (ID 90901331004) during the October 2023 X-ray outburst as presented in Table~\ref{tab:log}. The data reduction was carried out using the HEASoft v.6.32 and the calibration files of version 20231017. We followed the standard data processing routines where {\tt NUPIPELINE} and  {\tt NUPRODUCTS} tasks are executed to reprocess the unfiltered files in order to extract light curves and spectra. For source and background products, source and background regions were chosen using \texttt{ds9} software. The source region is centered on the source position, while the background region is chosen as far from the source position as possible. Both regions are circular in shape with radii of 180 arcsec and 90 arcsec for ID 90901331002 (ID 02 from now onwards) and 90901331004 (ID 04 from now onwards), respectively.

\begin{table}
	\centering
	\caption{Log of X-ray \& optical observations of IGR~J06074+2205}
	\label{tab:log}
	\begin{tabular}{lccc} 
		\hline

Obs. ID & Date of observation & MJD  &   Exposure time (s)    \\ [2 pt]

\hline 

\large{\textit{NuSTAR}}\\ [4 pt] 
90901331002 & 2023-10-08 & 60225 & 42704 \\
90901331004	& 2023-10-15 & 60232 & 40676 \\
\hline
\large{\textit{NICER}}\\ [4 pt] 
6204040101 & 2023-10-12 & 60229 & 163 \\
6204040102 & 2023-10-13 & 60230 & 2406 \\
6204040103 & 2023-10-14 & 60231 & 36 \\
6204040104 & 2023-10-15 & 60232 & 263 \\
6204040105 & 2023-10-17 & 60234 & 3497 \\
6204040106 & 2023-12-28 & 60306 & 10339 \\
6204040107 & 2023-12-29 & 60307 & 359 \\
6204040108 & 2023-12-29 & 60307 & 207 \\
6204040109 & 2024-01-01 & 60310 & 472 \\
6204040110 & 2024-01-05 & 60314 & 869 \\
6204040111 & 2024-01-06 & 60316 & 2532 \\
6204040112 & 2024-01-08 & 60317 & 572 \\
6204040113 & 2024-01-10 & 60319 & 1134   \\
\hline

\large{\textit{MIRO}} \\ [4 pt]
Epoch1 & 2022-11-18 & 59901  &  500 \\
Epoch2 & 2022-12-13 & 59926  & 400   \\
\hline
\large{\textit{IAO}} \\ [4 pt]
Epoch3 & 2023-10-22 & 60239 & 500 \\
Epoch4 &  2023-12-15 & 60293 & 500 \\
Epoch5 & 2024-02-03 & 60343 & 500\\
\hline
\end{tabular}
\footnotesize{ Note: The NICER exposure time corresponds to the net exposure time after applying SUNSHINE==0 filter.}
\end{table}

\subsection{NICER}
NICER, launched in June 2017 and installed on the International Space Station, is equipped with the X-ray Timing Instrument (XTI; \citealt{2016SPIE.9905E..1HG}) that is designed to operate in 0.2-12 keV energy range. The XTI comprises 56 X-ray concentrator optics, each paired with a silicon drift detector, providing non-imaging observations  \citep{2016SPIE.9905E..1IP}. These concentrator optics are comprised of 24 nested grazing-incidence gold-coated aluminum foil mirrors of parabolic shape. The XTI offers a high time resolution of approximately 100 ns (rms) and a spectral resolution of about 85 eV at 1 keV. Its field of view covers approximately 30 arcmin$^{2}$ in the sky. The effective area of NICER is approximately 1900 cm$^{2}$ at 1.5 keV, utilizing 52 active detectors. The XTI is divided into seven groups of eight focal plane modules (FPMs), with each group managed by a Measurement/Power Unit (MPU) slice. This arrangement enables independent operation and control of the FPMs within each group. 

Our study uses publicly available NICER data of IGR~J06074+2205 between MJD 60229.76 (12 October 2023) and MJD 60319.39 (10 January 2024). These data are stored under observation IDs 6204040xxx  with a net exposure time of $\approx$23 ks (see Table~\ref{tab:log}). We use {\tt nicerl2} pipeline available in \textsc{HEASoft} version 6.32 to process the unfiltered event data of NICER. The analysis is performed in the presence of gain and calibration database files of version \texttt{xti20221001}. We selected good time intervals (GTI) using the \texttt{nimaketime} tool, applying standard filtering criteria based on various factors: the South Atlantic Anomaly (SAA), an elevation angle exceeding $15^\circ$ from the Earth limb, a $>30^\circ$ offset from the bright Earth, and a pointing offset of 0.015$^\circ$, all on clean data post-{\tt nicerl2}. Additionally, we applied a magnetic cutoff rigidity of 4 GeV/c to filter out any potential high-energy charged particle background. Due to a known optical leak\footnote{\url{https://heasarc.gsfc.nasa.gov/docs/nicer/}} in XTI, we only considered night-side data in our analysis using SUNSHINE==0 filter. We also applied barycentric correction using the solar system ephemeris \texttt{JPL-DE430} to correct for Earth and satellite motion during observations. Subsequently, we extracted light curves and spectra from each observation using the \textsc{XSELECT} task. We utilized the {\tt nibackgen3C50} \citep{2022AJ....163..130R} tool to generate the background spectrum for each observation. Finally, we created a spectral response matrix and ancillary response files using the {\tt nicerrmf} and {\tt nicerarf} tasks.

\section{Optical observations \& data reduction}
\label{sec:3 optical reduc}
\subsection{MIRO}
We use the 1.2 m telescope of MIRO with the Mt. Abu Faint Object Spectrograph and Camera-Pathfinder (MFOSC-P) instrument \citep{2018SPIE10702E..4IS,srivastava2021design} mounted on the 1.2 m, f/13 telescope for two epochs of observations of IGR~J06074+2205 binary as presented in Table~\ref{tab:log}. The instrument is designed to provide seeing limited imaging in Bessel BVRI filters with a sampling rate of 3.3 pixels per arc-second over a 5.2$\times$5.2 arc-minute$^{2}$ field-of-view. The MFOSC-P is facilitated with three plane reflection gratings, named R2000, R1000, and R500, with 500, 300, and 150 lp/mm at wavelengths of $\approx$6500~\AA, 5500~\AA~ and 6000~\AA, respectively. R2000, R1000, and R500 represent the spectral resolutions (R=$\frac{\lambda}{\Delta\lambda}$) of 2000, 1000, and 500, respectively. These three modes provide a standard spectral coverage of $\approx$6000-7000~\AA, $\approx$4700-6650~\AA~ and $\approx$4500-8500~\AA, respectively. The spectroscopic observations are carried out using a 75 $\mu$m slit in R2000 mode, the highest resolution available for the MFOSC-P instrument.

The raw data of the MFOSC-P observations are reduced using the in-house developed data analysis routines in Python\footnote{ \url{https://www.python.org/}} with the assistance of open-source image processing libraries (ASTROPY\footnote{ \url{https://www.astropy.org/}}, etc.) (also refer to \citealt{2022MNRAS.510.4265K} for the analysis method). In the beginning, the bias is subtracted, cosmic rays are removed, and sky-subtracted images are created. Using a halogen lamp, the pixel-to-pixel efficiency variation is found to be less than 1\%; hence, no correction is applied. In the second step, the spectra of Neon and Xenon calibration lamps (taken after each source exposure) are used for the wavelength calibration of the data. For the instrument response correction, a spectro-photometric standard star from the European Southern Observatory (ESO) catalogue\footnote{\url{https://www.eso.org/sci/observing/tools/standards/spectra/okestandards_rev.html}}, G191B2B is used. The response function is calculated by dividing the observed continuum spectra of G191B2B with a blackbody curve corresponding to the standard star's effective temperature. The final spectra are created by applying the response function to the source spectra.

\subsection{IAO}
Three epochs of spectroscopic observations of IGR~J06074+2205 (see Table~\ref{tab:log}) are also carried out with the Hanle Faint Object Spectrograph and Camera (HFOSC) instrument mounted on the 2.01-m Himalayan Chandra Telescope (HCT) of  IAO, located at Hanle, Ladakh, India~\citep{2002ASPC..266..424C}. The spectra are taken with Grism~8 and 167~l slit setup, which covers a wavelength range of 5500–8350~\AA, providing an effective resolution of 2000 at H$\alpha$. Halogen lamp frames are taken for flat fielding the images. Data reduction is performed using self-developed data analysis routines in Python with the assistance of open-source image processing libraries (ASTROPY etc.) and standard IRAF tasks. In the first step, bias subtraction, cosmic-ray removal~\citep{2001PASP..113.1420V}, and flat-fielding are done for all frames. Then, one-dimensional spectra are extracted after removing the mean sky background taken from both sides of the spectrum. In the second step, we extract wavelength-calibrated spectra using FeNe lamp spectra. The instrument response function is generated following the procedure outlined for MFOSC-P/MIRO, with G191B2B as the standard star. Subsequently, the science-ready spectra are produced by applying this response function to the source spectra.

\begin{figure}
    \centering
    
    {\hspace*{-0.2cm}
    \includegraphics[trim={0 3cm 0 0},scale=0.45]{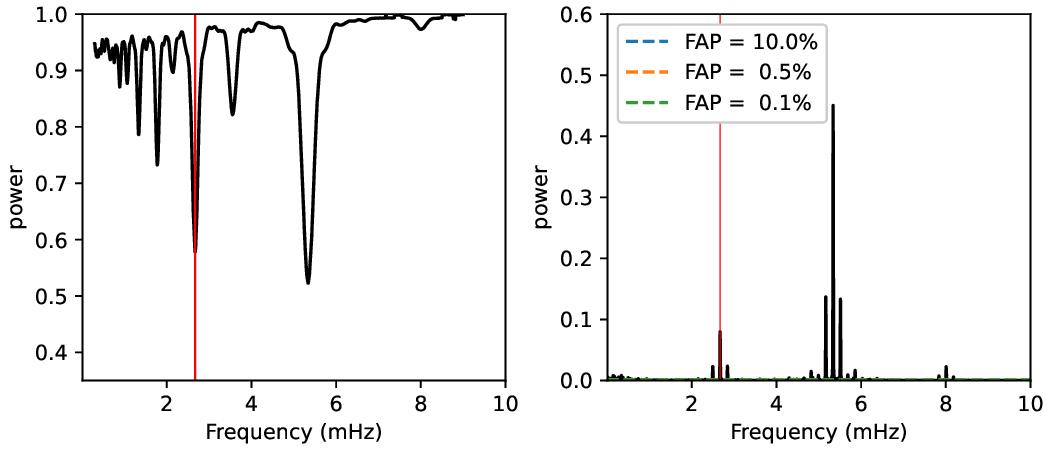}}
    \hspace*{-0.0cm}
    {\includegraphics[trim={0 2cm 0 0},scale=0.75]{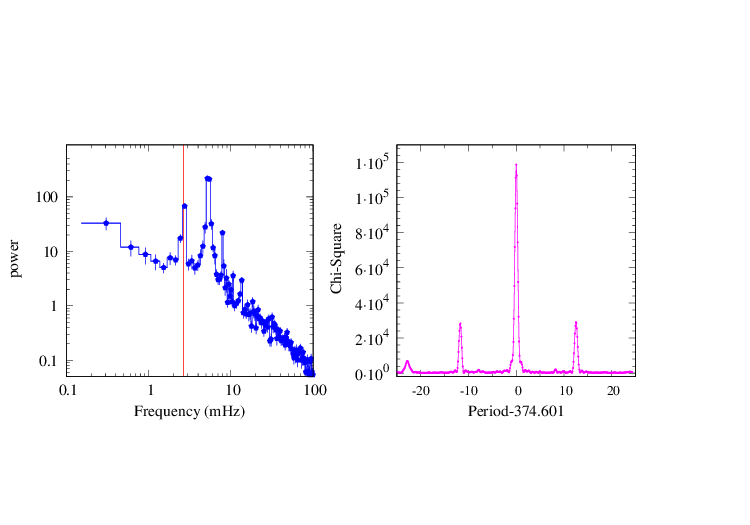}}
    \caption{Detection of pulsation at $\sim$2.66 mHz (374.60 s) for observation ID~02 using various techniques such as PDM (upper left), powspec (lower left), GLS periodogram (upper right), and efsearch (lower right) are shown. }
    \label{fig:pulse_detection}
\end{figure}

\begin{figure}
    \hspace*{-1.2cm}
    \includegraphics[trim={0 1.6cm 0 0}, scale=0.8]{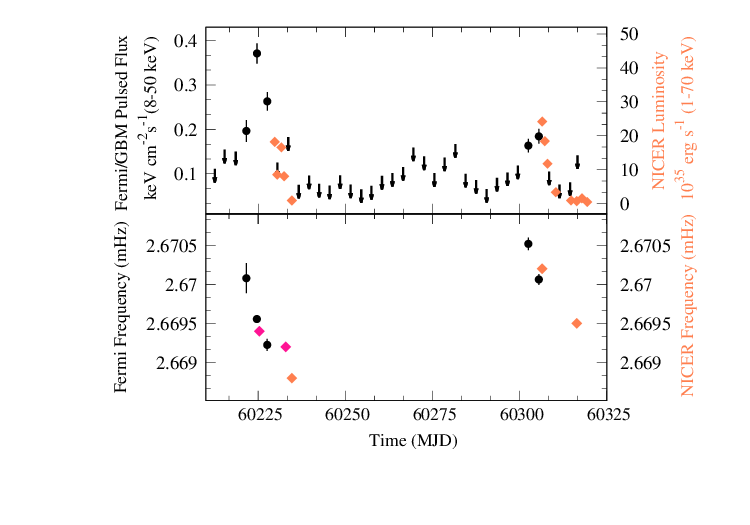}
    \caption{ Upper panel: Evolution of Fermi/GBM flux in the 8-50 keV band (black) and  NICER luminosity  in the 1-70 keV band (orange) over time. Bottom panel: Time evolution of the spin frequency of IGR~J06074+2205 measured using Fermi/GBM (black), NICER (orange), and NuSTAR (magenta) data.}
    \label{fig:spin_evolution}
\end{figure}

\begin{figure*}
    \centering
    \hspace*{1cm}
    \includegraphics[trim={0 1.2cm 0 0.5cm},scale=1.3]{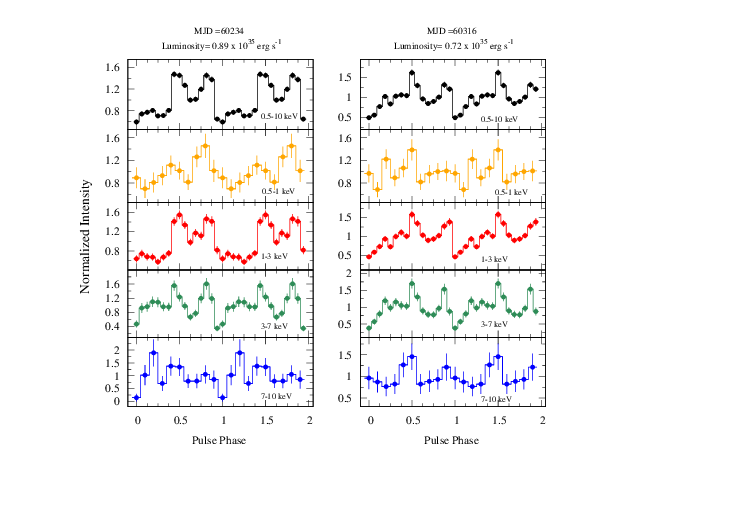}
    \caption{  Pulse profiles of the pulsar for the NICER observations IDs~105 (left panel) \& 111 (right panel), generated at energy ranges of 0.5-10 keV, 0.5-1 keV, 1-3 keV, 3-7 keV, and 7-10 keV are displayed from top to bottom, respectively. The luminosity and day of the observation (MJD) of each observation ID are provided at the top of the figure. }
    \label{fig:2idNICER_pp}
\end{figure*}

\section{X-ray timing analysis \& results}
\label{sec:xray_timing}
The X-ray timing analysis is conducted using data from the NuSTAR and NICER observations of the pulsar during its X-ray outbursts in October and December 2023 (see Table~\ref{tab:log}). The barycenter-corrected NuSTAR and NICER light curves are generated at a bin size of 10 ms in the energy range of 3-79 keV and 0.5-10 keV, respectively. At first, we search for pulsations in the light curves at a period close to 374 s~\citep{2018A&A...613A..52R} following the $\chi^{2}$-maximization technique \citep{1987A&A...180..275L} by using the \texttt{efsearch} task of \texttt{FTOOLS} package. The detection of pulsar period (period corresponds to highest $\chi^{2}$) for observation ID~02 using \texttt{efsearch} task is shown in the lower right panel of  Figure~\ref{fig:pulse_detection}. Secondly, we search for pulse period using the \texttt{powspec} tool from the \texttt{XRONOS} package. We employ the command of the \texttt{powspec norm=-2} to obtain white-noise subtracted averaged power density spectrum (PDS). Following this, the power is expressed in the units of (RMS/mean)$^{2}$/Hz. The obtained PDS is shown in the lower left panel of Figure~\ref{fig:pulse_detection}. Then, we apply the generalized Lomb-Scargle method (GLS; \citealt{2009A&A...496..577Z}) and Phase Dispersion Minimization (PDM; \citealt{1978ApJ...224..953S}) method to calculate the pulsation period of the neutron star. In Figure~\ref{fig:pulse_detection}, we show the pulse period search results for ID~02. For ID~02, the obtained pulse frequency is  $\approx$2.66 mHz (374.60 s). The obtained pulse frequency values are similar to all the methods. The frequency corresponds to the neutron star's pulsation or the fundamental frequency, and its corresponding power is marked as a vertical line. We observe that the power of 1st harmonic is more than the fundamental frequency, which is also the case for IGR~J06074+2205 during the 2017 observation ~\citep{2018A&A...613A..52R}.

 The spin periods obtained using the \texttt{efsearch} task are 374.60$\pm$0.30 s and 374.64$\pm$0.10 s for NuSTAR observation IDs 02 and 04, respectively. For NICER observation IDs 6204040105 (ID 105 from now onwards), 6204040106 (ID 106 from now onwards), and 6204040111 (ID 111 from now onwards), the estimated spin periods obtained using the \texttt{efsearch} task are 374.63$\pm$0.85 s, 374.54$\pm$0.35 s, and 374.57$\pm$0.35 s, respectively. The spin period values obtained from other methods are consistent with these measurements within the quoted uncertainties. It is possible to calculate the pulse period of the pulsar only in these 3 NICER observation IDs out of 13, which may be because of a lack of sufficient exposure times and a weak count rate. The error presented in each pulse period value is determined by fitting a Gaussian to the distribution of $\chi^{2}$ values obtained from the  $\chi^{2}$-maximization technique and considering 1-$\sigma$ standard deviation. We also use publicly available Fermi/GBM\footnote{\url{https://gammaray.nsstc.nasa.gov/gbm/science/pulsars/lightcurves/igrj06074.html}} spin frequency data \citep{2020ApJ...896...90M} for IGR~J06074+2205 to check for the spin frequency value and it's evolution with time and luminosity. The spin frequency evolution of the source with time obtained from the NuSTAR, NICER, and Fermi/GBM observations is shown in the bottom panel of Figure~\ref{fig:spin_evolution}. We present the evolution of 8-50 keV pulsed flux from Fermi/GBM and the source luminosity  in the 1-70 keV energy range from the NICER observations in the top panel of Figure~\ref{fig:spin_evolution}. The pulse frequency is found to decrease with time during the October \& December X-ray outbursts.

\begin{figure}
    \centering
    \hspace*{0.1cm}
    \includegraphics[trim={0 2cm 0 0.5cm},scale=0.45, angle=-90]{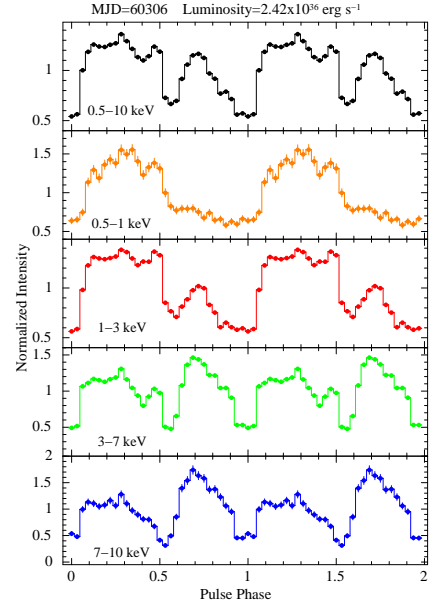}
    \caption{Pulse profiles are presented for NICER observation ID 106 generated at energy bands of 0.5-10 keV, 0.5-1 keV, 1-3 keV, 3-7 keV, and 7-10 keV and displayed from top to bottom, respectively. The luminosity and day of the observation (MJD) are provided at the top of the figure.  }
    \label{fig:niEreolved_pp}
\end{figure}

\begin{figure}
    \centering
    \includegraphics[trim={0.0cm 0cm 0.0cm 0.0cm},scale=0.5, angle=-90]{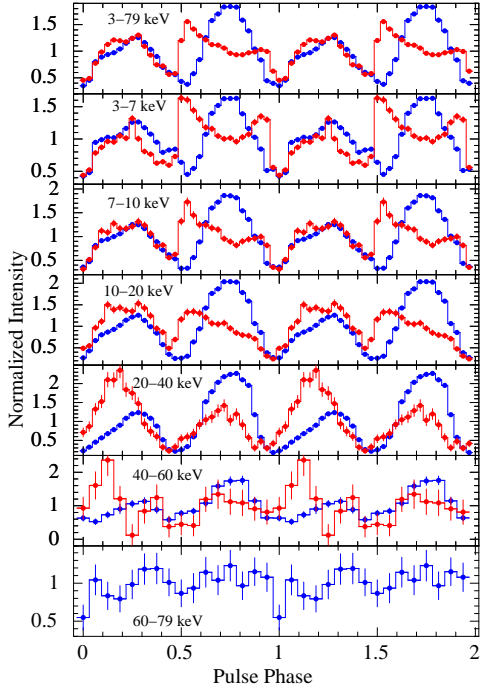}
    \caption{Energy-resolved pulse profiles are presented for NuSTAR observations ID~02 (blue) and ID~04 (red). The pulse profiles are obtained in the energy bands of 3-79 keV, 3-7 keV, 7-10 keV, 10-20 keV, 20-40 keV, 40-60 keV, and 60-79 keV displayed from top to bottom, respectively.}
    \label{fig:nuEreolved_pp}
\end{figure}

\begin{figure}
    \hspace*{-0.5cm}
    \includegraphics[trim={0 2.2cm 0 0.7cm},scale=1]{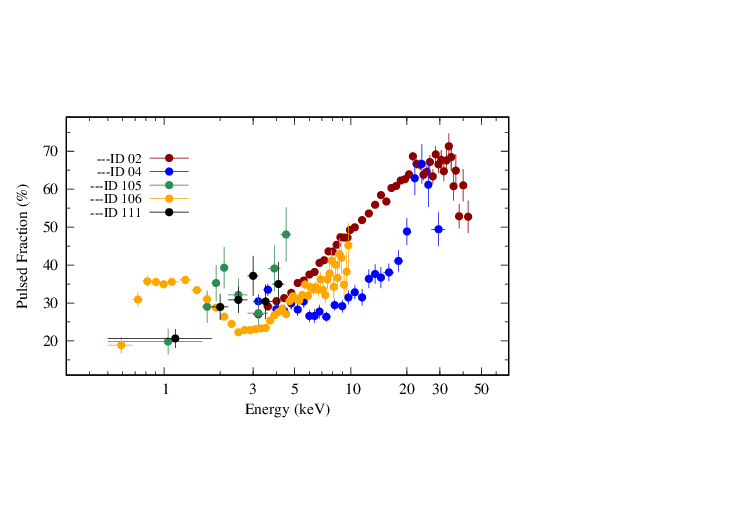}
    \caption{  Evolution of the pulsed fraction (PF) with energy calculated at SNR of 12 and 8 for NuSTAR (ID 02 \& 04) and NICER (ID 105, 106, and 111) data, respectively.}
    \label{fig:pf_energy}
\end{figure}

\begin{figure*}
    \hspace*{-1.2cm}
    \includegraphics[trim={0 0.0cm 0 0},scale=1.1]{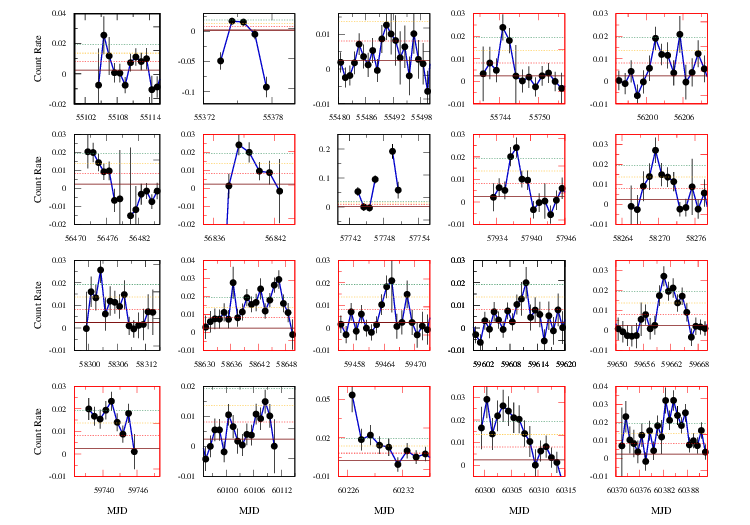}
    \caption{  Possible X-ray outbursts observed in IGR~J06074+2205 with MAXI/GSC in the 4-10 keV energy band. The brown line presents the mean count rate distribution observed in the source. The red, yellow, and green dotted lines mark the 1$\sigma$, 2$\sigma$, and 3$\sigma$ count rates above the mean. The mean and standard deviations are obtained by fitting a Gaussian to the histogram plot of count distribution as shown in the top panel of Figure~\ref{fig:maxi_distribution}. Panels with a red border contain outbursts that are chosen for further analysis. }
    \label{fig:outburst_candidate}
\end{figure*}
Subsequently, we analyze the energy-averaged and energy-resolved pulse profiles of the pulsar to investigate the geometry of emission and its energy dependence. Each light curve is folded at the corresponding spin period of the pulsar utilizing the \texttt{efold} task within the \texttt{FTOOLS} package.  In the beginning, we generated energy-averaged and energy-resolved pulse profiles for three NICER observation IDs to probe the emission geometry in the soft X-ray regime. The pulsar was relatively faint during IDs~105 \& 111 with 1-70 keV luminosities of ~8.9 and 7.2 $\times$10$^{34}$ erg s$^{-1}$ (estimated from spectral fitting, assuming a distance of 5.99 kpc; Section~\ref{sec:xray_spectral}), respectively. The pulse profiles of the pulsar for IDs~105 \& 111 are presented in Figure~\ref{fig:2idNICER_pp}. For ID~106, the pulse profiles are obtained during a relatively brighter state at a luminosity of 2.42$\times$10$^{36}$ erg s$^{-1}$ (see Figure~\ref{fig:niEreolved_pp}). 
 For the fainter observations (top panels of Figure~\ref{fig:2idNICER_pp}), the pulse profiles are double-peaked, with the two peaks appearing at phases $\sim$0.45 and $\sim$0.85. For ID~106, the pulse profile is also double-peaked with two peaks appearing at phases $\sim$0.3 (primary peak) and $\sim$0.7 (top panel of Figure~\ref{fig:niEreolved_pp}). Moreover, the primary peak also contains a minor dip at phase $\sim$0.4, unlike in the fainter observations.

We also generated energy-resolved pulse profiles in 0.5-1, 1-3, 3-7, and 7-10 keV energy bands for NICER observations (see Figures~\ref{fig:2idNICER_pp} \& \ref{fig:niEreolved_pp}). The pulse profiles from NICER observations IDs~105 \& 111, when the source was relatively faint, do not show significant variation with energy. However, the pulse profiles at brighter state exhibit notable energy-dependent changes. Below 1 keV, the profile is single-peaked, showing only the primary peak. However, with increase in energy, a secondary peak emerges in the 0.6–1.0 phase range and eventually becomes the dominant feature in the profile as reported by \citet{2024ATel16401....1N}.

To understand the behavior of the pulse profile across a broad energy regime, we used data from both the epochs of NuSTAR observations of IGR~J06074+2205 during the October 2023 outburst. For ID~02 and ID~04, the pulse profiles are generated at the luminosities of 5.56$\times 10^{36}$ and 4.66$\times 10^{35}$ erg s$^{-1}$ (in 1-70 keV band), respectively. The pulse profiles obtained in the 3-79 keV energy band are double-peaked and asymmetric at lower and higher luminosity state observation (1st panel of Figure~\ref{fig:nuEreolved_pp}). However, the pulse profile obtained at higher luminosity is relatively smoother than that at lower luminosity, which contains dip-like features in the 0.6--0.8 phase range. Further, we generated energy-resolved pulse profiles across energy bands of 3-7, 7-10, 10-20, 20-40, 40-60, and 60-79 keV, as depicted in Figure~\ref{fig:nuEreolved_pp}. The basic shape of the pulse profiles at higher luminosity did not vary much with energy. However, the peak in the phase range of 0.0-0.5 becomes dimmer, and the peak between phases 0.5 and 1.0  becomes brighter as the energy increases from 3 to 40 keV. Beyond 40 keV, the pulse profiles are coarsely binned due to low SNR. We observed that the difference between the strength of both peaks in these pulse profiles decreased. For ID~04, the basic shape of the pulse profiles varied significantly with energy. The peak in the 0.5-1.0 phase range harbors a deeper absorption feature in the 3-7 keV energy band, which diminished as the energy increased and vanished beyond 20 keV. Also, as the energy increases from 3 to 60 keV, the pulse peak in the 0.0-0.5 phase range evolved from the weaker to the dominant peak in the pulse profile. Overall, the pulse profiles are asymmetric and double-peaked, with differing shapes for both IDs below $\approx$20 keV~\citep{2024arXiv240217382R}.

Furthermore, we calculated the pulsed fraction (PF) of the pulse profiles to quantify the amount of pulsed emission from the source using the root mean square (RMS) method given by: 
\begin{equation}
   PF = \frac{(\sum_{i=1}^{N} (r_{i}-\overline{r})^{2}/N)^{1/2}}{\overline{r}}
\end{equation}
\label{eq:1}

Where, $r_{i}$ is the count rate in the $i$th phase bin of the pulse profile, $\overline{r}$ is the average count rate, and $N$ is the total number of phase bins. 
The PF variation with energy for NuSTAR observations (IDs~02 and~04) and NICER observations (IDs~105,~106, and~111) is shown in Figure~\ref{fig:pf_energy}. We calculated the PF from light curves in narrow energy ranges by following the technique of \citet{2023A&A...677A.103F}. As the energy resolution for NuSTAR is 400 eV at 10 keV and 900 eV at 68 keV~\citep{2013ApJ...770..103H}, the light curves are generated with energy bin spacings of 0.4 keV below 10 keV and 1.0 keV above 10 keV. Similarly, the light curves are generated with an energy bin of 0.09 keV below 1 keV and 0.2 keV beyond 1 keV for NICER observations. Energy-resolved pulse profiles are then created with 16 phase bins per period. Finally, the pulse profiles are combined to achieve an SNR of 12 for NuSTAR and 8 for NICER data, following \citet{2023A&A...677A.103F}. The SNR is calculated using the formulas suggested by \citet{2023A&A...677A.103F}:

\begin{equation}
    SNR= \frac{\Sigma|p_{i} - \Bar{p}|}{\sqrt{\Sigma(\sigma_{p_{i}})^{2}}}
\end{equation}

Where, p$_{i}$, $\sigma_{p_{i}}$ and $\Bar{p}$ are respectively the rate on the ith phase bin, uncertainty, and the average rate of the pulse profile. Then, we calculate the PF using the RMS method (Equation~1).

 For ID~02, between 3 and 35 keV range, the PF value increases monotonically from $\approx$25\% to $\approx$70\%. Beyond 35 keV, it decreases sharply from $\approx$70\% to $\approx$50\% at 45 keV. For ID~04, the PF remains flat around 30\% between 3 and 8 keV and then increases to $\approx$70\% between 8 and 25 keV. Again, it decreases sharply from $\approx$70\% to $\approx$50\% between 25 and 30 keV. At higher energies, the PF is not calculated as it did not achieve the prescribed SNR.

 For NICER observations, we observe a distinct PF evolution with energy. For ID~105, the PF rises from $\approx$20\% to 40\% between 0.5 and 2 keV, then decreases to $\approx$25\% between 2 and 3 keV, and then increases again to $\approx$50\% between 3 and 5 keV. For ID~111, the PF varied between $\approx$20\% to 35\% as energy increased from 0.5 to 5 keV. For ID~106, the PF shows a unique broad hump-like feature between 0.5 and 3 keV, varying between 20\% and 40\%, followed by a steady increase from $\approx$20\% to 40\% in the 3-10 keV energy band.

\subsection{Orbital period determination of IGR~J06074+2205}
\label{sub:orbital period}
\begin{figure}
    \centering
    \includegraphics[trim={0 0.0cm 0 0},scale=0.4]{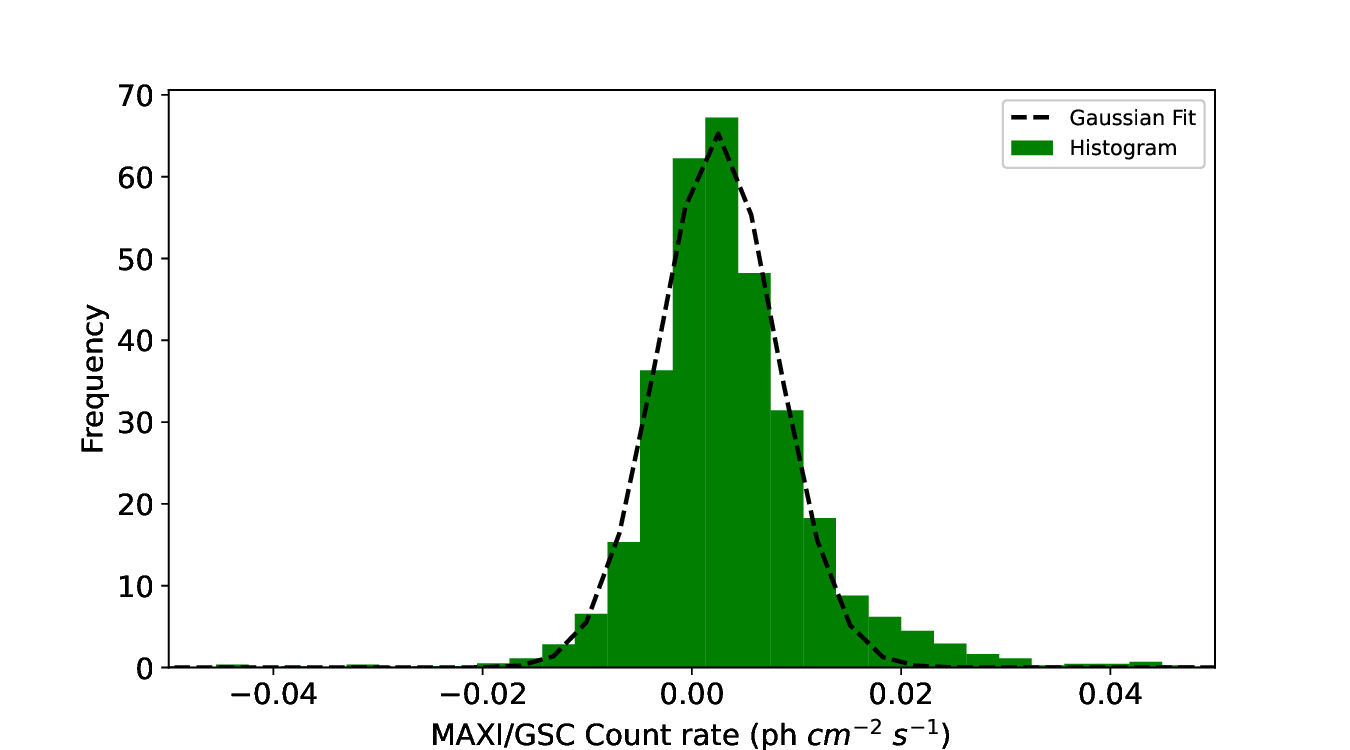}\vspace*{-1cm}
\hspace*{-4cm}\includegraphics[trim={0 0.0cm 6.0cm 0},scale=0.72]{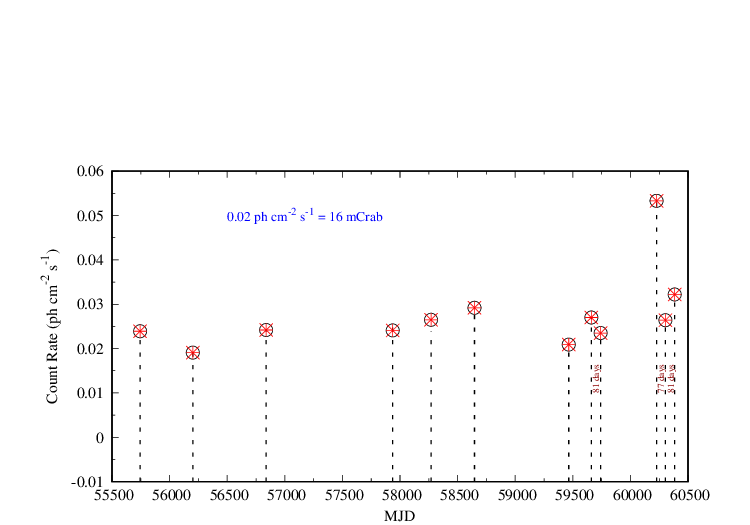}
    \caption{Upper Panel: Histogram distribution of MAXI/GSC count rate in 4-10 keV band. A Gaussian function is fitted to obtain the mean and standard deviation of the count rates. Bottom panel: Occurrence times of 12 chosen X-ray outbursts and their corresponding peak intensities are shown. The time differences between closer outbursts are also annotated.}
    \label{fig:maxi_distribution}
\end{figure}

Though we have some information about the Be star and neutron star in the Be/X-ray binary IGR~J06074+2205, the orbital parameters of the binary are yet to be determined.  \citet{2023ATel16351....1M} suggested the orbital period of the binary to be 80$\pm$2 or 80/n (n=1,2,3...) by analyzing four possible X-ray outbursts in MAXI/GSC light curve. In this work, we conducted a comprehensive analysis by studying 20 possible X-ray outbursts. We started by looking at the long-term light curve of IGR~J06074+2205, utilizing data from MAXI/GSC (4-10 keV range) as it provides frequent monitoring of the source compared to Fermi/GBM and Swift/BAT. Our approach is to find signatures of periodic X-ray enhancements in the X-ray light curve of IGR~J06074+2205, which is assumed to occur due to mass accretion by a neutron star during its periastron passage. IGR~J06074+2205 has rarely shown any remarkable normal or giant X-ray outbursts since its discovery in 2003. However, we attempted to find any periodic enhancement in the light curve. We initially applied two robust techniques, the GLS periodogram and the PDM method. Both techniques exhibit strong periodicity at $\approx$72 days using MAXI. However, this periodicity is known to arise due to the precession period of the International Space Station \citep{2022ATel15614....1C}. Then, we manually try to identify the outburst-like features in the long-term light curve. The possible 20 X-ray outbursts-like features from IGR~J06074+2205 are selected using MAXI/GSC 4-10 keV band long-term light curve (see Figure~\ref{fig:outburst_candidate}). We use the 4-10 keV light curve instead of the light curve in the 2-20 keV range to avoid flare-like profiles for unexpected background increase or dip-like structure due to the shadow of solar panels, which are more prominent in 2-4 keV\footnote{\url{http://maxi.riken.jp/top/readme.html}}. Then, from the sample of 20 X-ray outburst-like features, we select 12 features for further analysis. These selected outburst-like features are marked in red color in Figure~\ref{fig:outburst_candidate}. These 12 outbursts (outburst-like features) are selected considering the peak intensity as 3$\sigma$ or above from the mean value and do not show sudden rise and decay of intensity. The day of occurrence (MJD) and corresponding X-ray intensities of these 12 outbursts are plotted in the bottom panel of Figure~\ref{fig:maxi_distribution} to give a clear picture of the occurrence time of the outbursts. We find that the minimum difference between two consecutive outbursts is approximately 80 days.  This result is consistent with the finding of \citet{2023ATel16351....1M}.

\section{ X-ray spectral analysis \& results}
\label{sec:xray_spectral}
The X-ray spectral analysis of 13 NICER observations and 2 NuSTAR observations of IGR~J06074+2205, during October and December 2023 outbursts is performed using XSPEC v-12.13.1 \citep{1996ASPC..101...17A} package. We initiate spectral analysis utilizing 13 NICER observations of the pulsar. The spectral analysis is restricted to the 1-7 keV energy band as in some observations, data beyond 7 KeV is background dominated. Spectra are binned to ensure a minimum of 20 counts per energy bin, facilitating the application of \texttt{Chi-Square} statistics in our analysis. For assessing the line-of-sight X-ray absorption, we utilize the \texttt{wilm} abundance table \citep{2000ApJ...542..914W}  alongside \texttt{Vern} photo-ionization cross section \citep{1996ApJ...465..487V}. As recommended by the instrument team, a systematic uncertainty of 1.5\% is incorporated. The spectra are effectively fitted with an absorbed power law (\texttt{TBabs*powerlaw}) model.

\begin{figure}
    \centering
    \hspace*{1.0cm}
    \includegraphics[trim={0 1cm 0 0},scale=0.9]{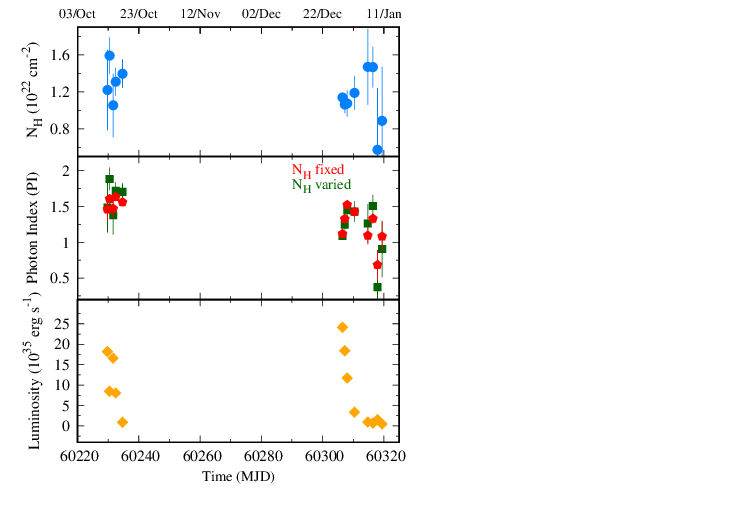}
    \caption{Temporal evolution of best-fitted spectral parameters N$_{H}$ (top panel), Photon Index (middle panel), Luminosity (bottom panel) obtained from the spectral fitting of the NICER observations.}
    \label{fig:NICER_spec}
\end{figure}

\begin{figure}
    \centering
    \includegraphics[trim={0 1cm 0 0},scale=0.7]{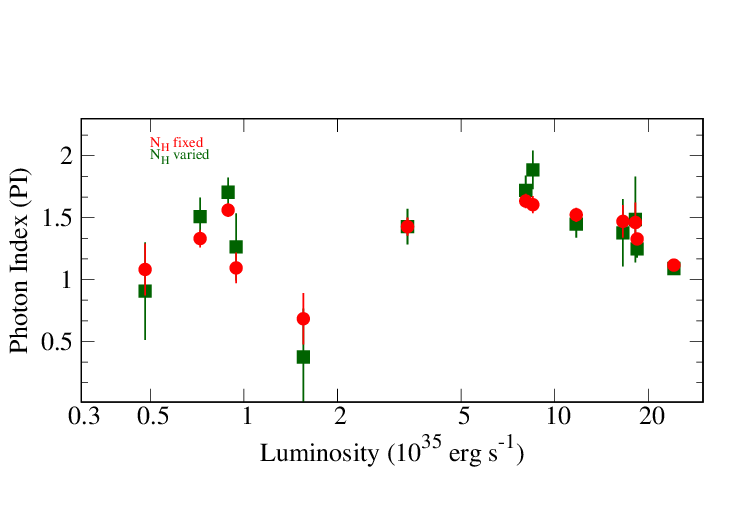}
    \caption{Correlation between the photon index (PI) and luminosity derived from the spectral fitting of NICER observations.}
    \label{fig:PIvsLUM}
\end{figure}

We observe a minimal variation in the value of equivalent hydrogen column density ($N_{H}$). The variations are within the errors. Therefore, we refit the spectra using an absorbed power law model by fixing the value of $N_{H}$ at the average value of 1.187$\times$ 10$^{22}$ cm$^{-2}$  \citep{1990ARA&A..28..215D, 2005A&A...440..775K, 2016A&A...594A.116H}. The temporal evolution of the best-fitted parameter values is shown in Figure~\ref{fig:NICER_spec}. The uncertainties on the parameters are calculated within the 90\% confidence range.  The luminosity for both NICER and NuSTAR observations is computed within the 1-70 keV energy range to maintain coherence and enable a consistent comparison of source properties, considering the source distance of 5.99 kpc \citep{2022A&A...665A..69F}.  The values of  photon index (PI) with and without fixed $N_{H}$ exhibit similar variation. We noticed that the average value of the PI obtained during the October outburst is relatively higher than the December outburst. The bottom panel of Figure~\ref{fig:NICER_spec} illustrates the luminosity variation, suggesting NICER observed the source during the declining phases of both the X-ray outbursts. Furthermore, to examine the relationship between the PI and luminosity, we plot them against each other in Figure~\ref{fig:PIvsLUM}.  We observe that the PI values fluctuated between 1 and 2 without exhibiting any particular pattern except for data points between 1-2$\times 10^{35}$ erg s$^{-1}$, where a sudden decrease in PI value is observed. It is challenging to draw conclusions from this due to large error bars.

\begin{figure}
\centering
\hspace*{-0.6cm}\includegraphics[trim={0 1cm 0 1.5cm},scale=0.88]{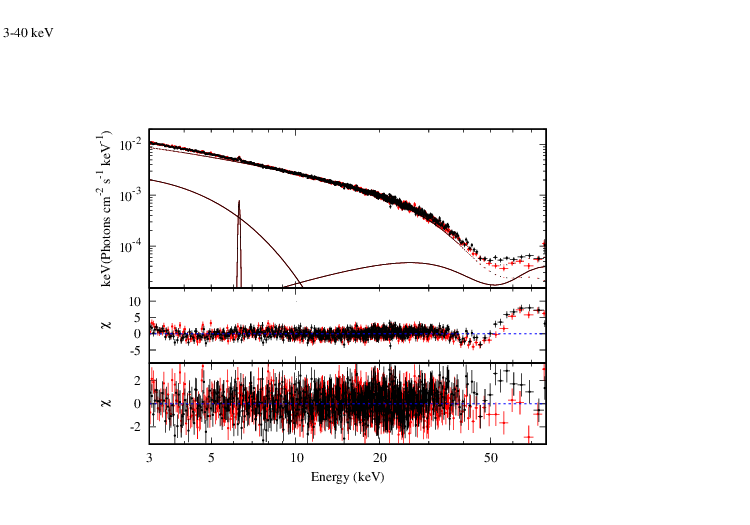}
\caption{The spectrum with best-fit model comprising an NPEX continuum model with a blackbody component, an iron emission line, and a Gaussian absorption component for CRSF, is shown. The middle and bottom panels show the residuals without and with the CRSF component in the spectral model, respectively.}
\label{fig:gabs_detection}
\end{figure}

\begin{figure}
    \hspace*{-0.2cm}
    \includegraphics[scale=0.35, angle=-90]{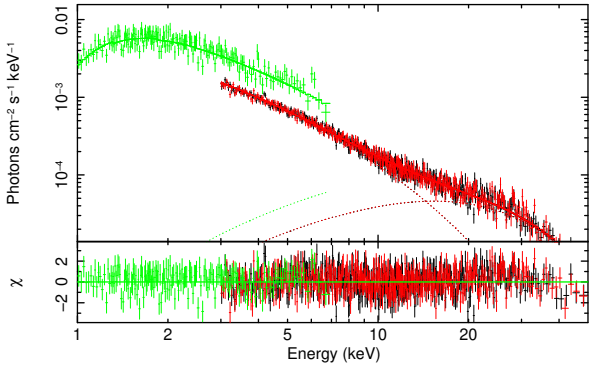}
    \caption{Spectra of IGR~J06074+2205 in the 1–50 keV range acquired from both NuSTAR and the NICER observations on 60232 MJD. The spectra are simultaneously fitted with the NPEX continuum model. The bottom panel illustrates the residuals derived from the best-fitting model.}
    \label{fig:npex_nicer_nustar}
\end{figure}

Subsequently, we conducted a broadband spectral analysis of two NuSTAR observations to investigate the hard X-ray emission properties and the presence of cyclotron resonance scattering feature (CRSF) in the pulsar spectrum.  In the first step, we compared the observed (source+background) and background spectra for both observations. For ID~02, It is noticed that the background and the observed spectra are comparable beyond $\sim$70 keV and $\sim$40 keV for observations ID~02 and ID~04, respectively (Figure~\ref{fig:id_04_source_back}). We confirmed this by examining the net observed (source+background) and background count rates in different energy bands (Table~\ref{tab:src_back_count}). For ID~02, the observed count rate remained above the 3$\sigma$ background variation up to 79 keV, whereas for ID~04, this is possible below 50 keV. Therefore, spectral analysis is performed in the 3-79 keV and 3-50 keV bands for IDs~02 and 04, respectively.

Initially, fitting the 3-79 keV spectrum of ID~02 with an absorbed power-law model revealed positive residuals near 6.4 keV and negative residuals near 50 keV. Positive residuals near 6.4 keV are modeled with a Gaussian component. Then, we use the Gaussian absorption line (\texttt{GABS}) model to account for the residuals near 50 keV. The inclusion of these components fits the spectra well and improves the $\chi^{2}$ value significantly.  The Gaussian absorption component near 50 keV is interpreted as the CRSF, consistent with findings of  \citet{2024arXiv240217382R} and ~\citet{2024JHEAp..42..129T}. We confirmed the presence of this absorption feature in the spectra using various continuum models, such as \texttt{CUTOFFPL}, \texttt{HIGHECUT}, \texttt{FDCUT}, and \texttt{NPEX}, all yielding comparable reduced~$\chi^{2}$ values. The spectral fitting results are detailed in Table~\ref{tab:specfit002}. The top panel of Figure~\ref{fig:gabs_detection} shows the source spectra and the best-fit model consisting of the NPEX continuum model \citep{1999ApJ...525..978M}, the Galactic absorption component, a blackbody, a Gaussian function for the 6.4 keV Fe emission line, and a Gaussian absorption component (GABS) for the CRSF.  The middle and bottom panels of the figure show the residuals obtained after fitting the spectra with the above model without and with the GABS component. From Table~\ref{tab:specfit002}, we observe that the absorption line parameters change depending on the choice of the continuum model. We also fitted the NuSTAR spectrum in 3-40 keV and 3-79 keV ranges to examine the effect of GABS on the continuum parameters (Table~\ref{tab:specfit002}). We found that the cutoff energy changed significantly when the GABS component was added to the model fitted in the 3-79 keV range. 

For ID~04, spectral fitting is limited to the 3-50 keV range as the data beyond 50 keV is background-dominated (Figure~\ref{fig:id_04_source_back}).  The 3-50 keV range spectrum is fitted well with the \texttt{NPEX} continuum model. We also carried out simultaneous NICER and NuSTAR spectral fitting using \texttt{NPEX} model in 1-50 keV band. The best-fitted spectral parameters are presented in Table~\ref{tab:specfit_nicer_nustar}. The best-fit model and corresponding residuals are shown in Figure~\ref{fig:npex_nicer_nustar}.

 For ID~04, \citet{2024arXiv240217382R} conducted spectral fitting in the 3-78 keV range and reported a 10 keV absorption-like feature, although they noted poor statistical significance beyond 40 keV. In contrast, \citet{2024JHEAp..42..129T} focused on the 3-50 keV range due to background dominance beyond 50 keV and found evidence for a blackbody component with kT $\sim$ 1 keV. We also tested various empirical models, and our analysis indicates that the detection of these features is highly dependent on the chosen continuum model, with the NPEX model offering the best fit in the 1-50 keV range.

\begin{figure*}
    \centering
    \includegraphics[trim={0 0.5cm 0 1cm},scale=1.2]{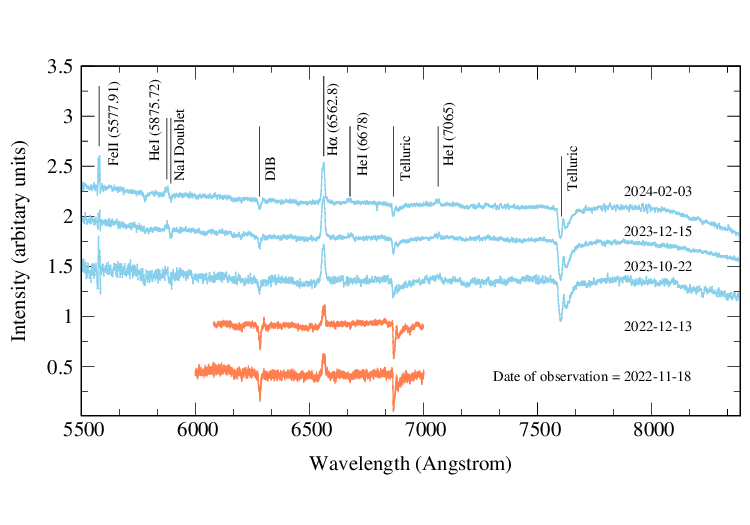}
    \caption{Evolution of the optical spectrum of Be/X-ray binary IGR~J06074+2205 as seen with the MFOSC-P (orange) \& HFOSC (blue) instruments. The observation dates are annotated on each spectrum.  The spectra are plotted with certain offsets for clarity.}
    \label{fig:optical_spectra}
\end{figure*}

\begin{figure*}
\begin{tabular}{cc}
\hspace{1cm}
\includegraphics[trim={0 0.7cm 0 0.7cm},scale=0.35]{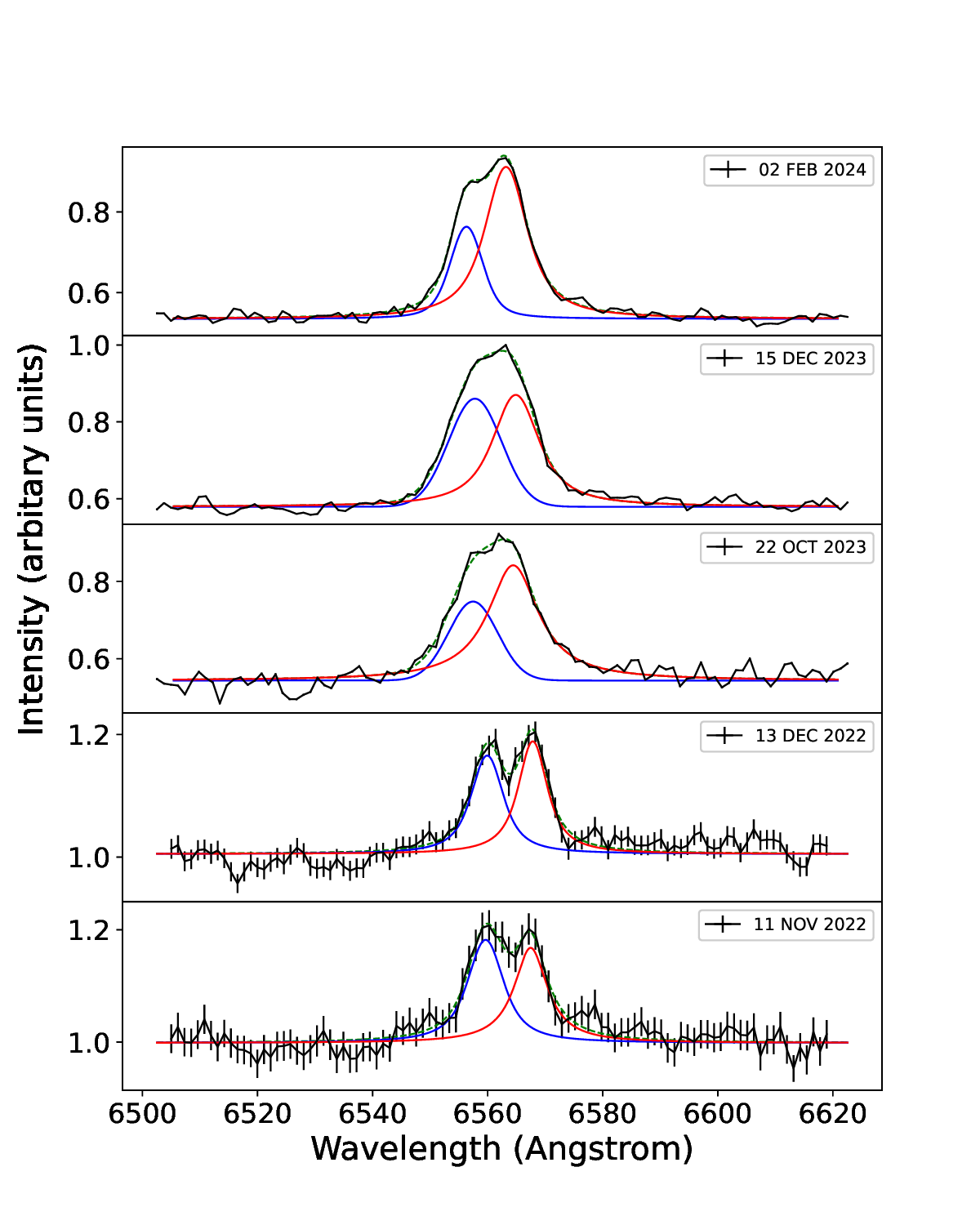} & \hspace{-1cm}
\includegraphics[trim={0 0.7cm 0 0.7cm},scale=0.35]{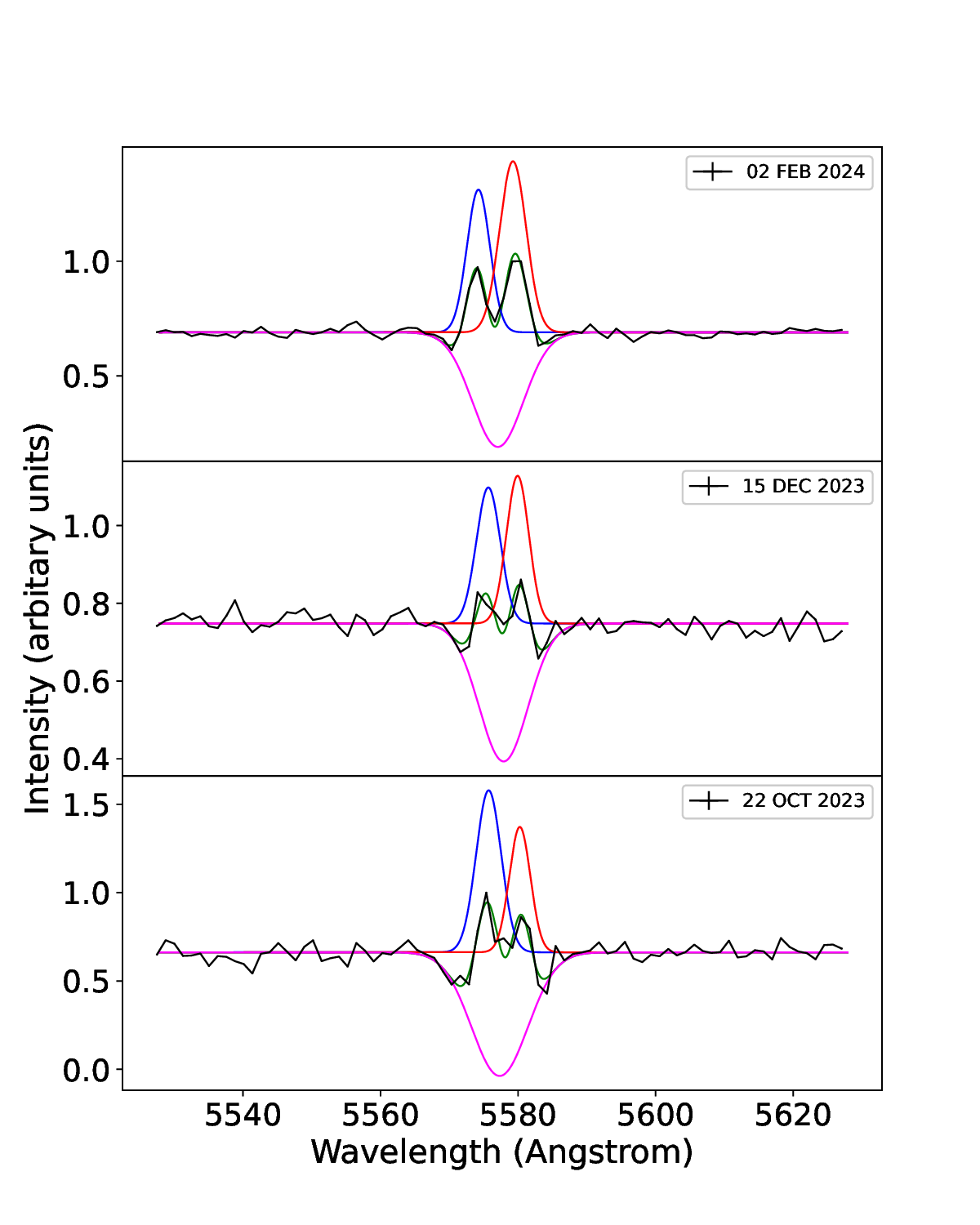}
\end{tabular}
\caption{Left: The observed H$\alpha$ line profiles (solid black lines) of IGR~J0674+2205 during our observations and corresponding best-fitted Voigt functions (dashed green lines). Right: Observed FeII line profiles (solid black line) of IGR~J0674+2205  and corresponding best-fitted Gaussian functions (dashed green line). The observation dates are quoted on the right of the corresponding profiles. The red and blue color profiles correspond to the red and blue shifted components of emission lines, respectively. }
\label{Fig:halpha_FeII_fitting}
\end{figure*}

\section{Optical spectroscopy \& results}
\label{sec:optical analysis}

We carry out optical spectroscopic analysis using five epochs of observations taken between 2022 \& 2024 with  MFOSC-P/MIRO \& HFOSC/IAO  at wavelength ranges of 6000-7000~\AA~and 5500-8350~\AA, respectively. The spectra from these observations are presented in Figure~\ref{fig:optical_spectra}. This figure shows the normalized spectrum for each observation epoch with certain offsets for clarity. The bottom two and top three spectra are obtained using MFOSC-P and HFOSC instruments, respectively. The date of each observation is also annotated in the figure.

In the optical spectra, we observe the presence of both emission and absorption lines. The prominent absorption lines are Diffuse Interstellar Band (DIB) and atmospheric telluric features. The DIBs are absorption features that arise when photons travel through significant column densities in the interstellar medium. We detect a DIB clearly at 6283.86~\AA~ \citep{1995ARA&A..33...19H}. We also observe a blended interstellar sodium doublet D1 and D2 in our spectra in absorption form \citep{1997ApJ...477..209K}. The telluric features arise because of the absorption of photons by the molecules in the Earth's atmosphere. These telluric features superimpose on the stellar spectra. Two most prominent telluric features such as O$_{2}$ B-band at $\lambda \sim 6870 $ \AA~ and O$_{2}$ A-band at $\lambda \sim 7605 $ \AA ~(\citealt{2019AJ....157..156A} \& reference therein) are present in our spectra.  A strong H$\alpha$~(6562.8~\AA) emission line is detected in the MOFOSC-P and HFOSC spectra of IGR~J06070+2205 (Figure~\ref{fig:optical_spectra}). We also detect another emission line at 5577.91~\AA~ in the HFOSC spectra due to its wider wavelength coverage towards the bluer side. In addition to that, we observe a weak signature of HeI (5875.72~\AA), HeI (6678~\AA), and HeI (7065~\AA) emission lines in the spectra taken from HFOSC on 15 December 2023 and 03 February 2024.

 To examine the structure and characteristics of the H$\alpha$ line profile, we focused on plotting only the portion of the optical spectra that covers the H$\alpha$ emission line region. The evolution of the H$\alpha$~(6562.8~\AA) emission line profile during all our optical observations is shown in the left part of Figure~\ref{Fig:halpha_FeII_fitting}. We calculate the equivalent width ($EW$) of the H$\alpha$ emission lines in the spectra from all epochs of our observations. The $EW$ signifies the strength of the emission/absorption lines.  It is a measure of the area of emission or absorption line in a wavelength vs intensity plot and is calculated by the following method:\newline \vspace{0.2cm}
 
 {\Large{ $EW$ = $\sum (1- \frac{F(\lambda)}{F_{c}(\lambda)})\delta \lambda$}} \newline \vspace{0.2cm}
 
where, F($\lambda$) represents the total flux consisting of contributions from the line and the continuum of the source spectrum, and F$_{c}(\lambda)$ represents only the underlying continuum flux at wavelength $\lambda$. The obtained $EW$ are presented in Table~\ref{tab:optical parameters}. The error in the value of $EW$ is calculated by the propagating error in the F$_{c}(\lambda)$.

 As shown in Figure~\ref{Fig:halpha_FeII_fitting}, the  H$\alpha$ line evolves from a double-peaked profile, with both peaks having similar intensities to a single peak dominated shape throughout our observations. The double-peaked H$\alpha$ line on 11 November 2022 has a relatively stronger blue-shifted peak than the red-shifted peak. However, the line profile on 13 December 2022 is the opposite, with a stronger red-shifted peak. After about one year, the H$\alpha$ line profile is dominated by the red-shifted peak. We fit all the H$\alpha$ emission lines with Voigt profiles to derive essential quantities like $V/R$ and peak difference ($\Delta$$\lambda$ or $\Delta$V) between the red and blue-shifted peaks. The $V/R$ value gives an idea of the contribution of the blue-shifted peak with respect to the red-shifted peak and vice versa. The peak difference information helps us to calculate the H$\alpha$ emitting region or the size of the circumstellar disc around the Be star. Initially, we try to fit the emission line profiles with Gaussian functions. However, the Gaussian functions do not fit well in the wing regions. The original and fitted line profiles are shown in black and green colors, respectively, in the left part of Figure~\ref{Fig:halpha_FeII_fitting}. The red and blue components of the lines are shown with red and blue colors, respectively, in the figure.

\citet{1972ApJ...171..549H} demonstrate that the size of the H$\alpha$ line emitting region of the circumstellar disc can be estimated by using the peak separation ($\Delta V$) of the double-peaked H$\alpha$ emission line, assuming Keplerian velocity distribution of matter in the disc. The velocity separation between the red and blue-shifted components of the emission lines can be used to calculate the radius of the emitting region \citep{1972ApJ...171..549H}. The relationship between the peak separation and the rotational velocity of the gas particles at the emitting region is given by $\Delta V$= 2$V_{rot}\sin i$. Following this, the size of the H$\alpha$ emitting region can be derived by :\newline \vspace{0.2cm} 
\begin{equation}
R_{d}= (2V \sin i / \Delta V)^{j} \epsilon R_{\ast} 
\end{equation}

where, $j$=2 for Keplerian rotation, $R_{\ast}$ is the Be star radius, and $\epsilon$ is a dimensionless parameter that considers several effects that would over-estimate the disc radius ($\epsilon$=0.9$\pm$0.1; \citealt{2013A&A...559A..87Z}). The value of $Vsini$ used in the calculation is 260 km s$^{-1}$ \citep{2010A&A...522A.107R}. We also calculate the full width at half maximum ($FWHM$), the width of the line at half of the peak flux levels of the emission or absorption line. The $FWHM$ can be expressed in the velocity unit to estimate the radial velocity of the H$\alpha$ emitting particles in the disc with respect to the observer. In this work, the $FWHM$ is calculated by fitting a single Gaussian function over the entire H$\alpha$ emission line using a Python routine that fits the overall line shape well. The values of the calculated parameters are given in Table~\ref{tab:optical parameters}. The values of $\Delta V$ and FWHM are found to vary within errors between November 2022 and February 2024 and remain at around 343~km~s$^{-1}$ and 700~km~s$^{-1}$, respectively. From Table~\ref{tab:optical parameters}, it can be noticed that the value of the EW increased with time from $\approx$2.5 to 11~\AA. The disc radius varies within errors and remains at around 2$R_*$. The errors in the values of $\Delta V$ and disc radius are found to be large for observations on 22 October 2023 and 15 December 2023 because of the propagation of larger error by $\Delta$V during the disc radius error calculation.

\begin{table*}
\caption{ Equivalent width, $\Delta V$, $FWHM$, Disc radius, and other spectral parameters of H$\alpha$ emission line. }
 \label{tab:optical parameters}
\centering
\setlength{\tabcolsep}{2pt}
\begin{tabular}{ cccccc@{\vline} ccccc }
\hline
 &                   & &H$\alpha$ (6562.8~\AA)   &           & &        &           &   &  FeII (5577.91~\AA)         &               \\
\hline 

 Date of&  $EW$ & V/R &$\Delta V$ & $FWHM$  &Disc Radius~ &  ~~$EW$ & V/R &$\Delta V$ & $FWHM$  &  Disc Radius  \\ [4 pt]
 observation  & (\AA) & & (km/s) & (km/s)  & ($R_*$) & ~~(\AA) & & (km/s) & (km/s)  & ($R_*$) \\ [4 pt]
\hline 
2022 Nov 18 & -2.51$\pm$0.20 &1.08$\pm$0.33  & 358$\pm$24 & 735$\pm$40 & 1.89$\pm$0.25& -& -&- & -&-  \\[6 pt]
2022 Dec 13 & -1.56$\pm$0.18&0.92$\pm$0.27  & 362$\pm$19 & 686$\pm$44 & 1.85$\pm$0.20& - & -& -& -&- \\[6 pt]
2023 Oct 22 & -9.83$\pm$0.87&0.47$\pm$0.27  & 332$\pm$96 & 729$\pm$23 & 2.22$\pm$1.29& ~~1.03$\pm$0.98 & 1.53$\pm$0.48 & 244$\pm$21 & 374$\pm$138&- \\[6 pt]
2023 Dec 15 & -10.33$\pm$0.38&0.81$\pm$0.21  & 336$\pm$50 & 700$\pm$13 & 2.15$\pm$0.63& ~~-2.86$\pm$0.44 &0.98$\pm$0.50 & 229$\pm$43 & 314$\pm$159& 4.64$\pm$1.75 \\[6 pt]
2024 Feb 03 & -11.21$\pm$0.52&0.41$\pm$0.06   & 326$\pm$16 & 648$\pm$14 & 2.28$\pm$0.22& ~~-4.42$\pm$0.40 &0.71$\pm$0.23 & 270$\pm$39& 433$\pm$58& 3.33$\pm$0.95\\[6 pt]

\hline
\end{tabular}
\end{table*}

We also detected another emission line at 5577.91~\AA~ and identified the line as FeII emission line by looking at NIST Atomic Spectra Database\footnote{\url{https://physics.nist.gov/PhysRefData/ASD/lines_form.html}}. Other atomic lines close to this wavelength are also present. However, we consider this  FeII line as the FeII lines are usually observed in the spectrum of Be stars (\citealt{2006A&A...460..821A} \& references therein). In our observations, the FeII line is detected only in the HFOSC spectra and exhibits dynamic behavior over time (Figure~\ref{Fig:halpha_FeII_fitting}). The FeII line obtained on 10 October 2023 and 15 December 2023 shows a weak emission shape along with significant dips towards the wings. These profiles can be termed as emission above absorption type double-peaked profiles. The FeII emission lines are double-peaked in shape and evolve from a slightly blue-shift-dominated profile to a red-shift-dominated profile between 10 October 2023 and 15 December 2023. The FeII line observed on 3 February 2024 exhibits a red-shift-dominated shape. We calculate the $V/R$ value and peak velocity separation ($\Delta$V) for the FeII emission line. At first, we fit the two Gaussian models for the red- and blue-shifted peaks. However, the FeII lines observed on 10 October 2023 and 15 December 2023 are not fitted well because of the strong absorption dips close to the wings due to the photospheric absorption of the central star. Then, we add two Gaussians for the blue and red-shifted emission components above another Gaussian absorption component. The addition of the Gaussian absorption component fits the line profile reasonably well (Figure~\ref{Fig:halpha_FeII_fitting}). The disc parameters derived using best-fitted spectral parameters are shown in Table~\ref{tab:optical parameters}. It can be seen that the EW increases from $\approx$1~\AA~to $\approx$4.5~\AA~ with time, and the $\Delta$V and FWHM, on average, remain around 250 \& 380 km s$^{-1}$, respectively. The radius of the FeII emitting region remains at around 4$R_*$.

\section{Discussion}
\label{discussion}

We carried out X-ray and optical studies of the Be/X-ray binary system IGR~J06074+2205. X-ray studies are performed during the October 2023 outburst using two NuSTAR observations and thirteen NICER observations during the October and December 2023 X-ray outbursts.  Optical spectroscopy was carried out using five epochs of observations from MIRO and IAO between 2022 and 2024, covering the phases of before, during, and after 2023 X-ray outbursts. These combined efforts aim to provide an understanding of the accretion emission from the neutron star and the evolution of Be decretion disc in the system.

We detected periodic signals of 374.6 s in NuSTAR and three NICER light curves, as detailed in Section~\ref{sec:xray_timing}. Additionally, we explored pulsation information from the Fermi/GBM database, the results of which are depicted in Figure~\ref{fig:spin_evolution}. The bottom panel of Figure~\ref{fig:spin_evolution} shows a decreasing trend in the pulsar spin frequency during the October and December 2023 X-ray outbursts. A decrease in spin frequency or increase in spin period during X-ray outbursts is usually not observed in accretion-powered X-ray pulsars.  Additionally, we calculated the rate of change of the spin period during October and December X-ray outbursts using the spin period values obtained from NuSTAR, NICER, and Fermi/GBM data. We found that the spin period is increased at the rate of 0.024$\pm$0.001 and 0.02$\pm$0.01 s day$^{-1}$ during October and December 2023 X-ray outbursts, respectively. This increase in the spin period may be attributed to the binary motion of the neutron star. However, confirmation of this is hindered by insufficient detailed information regarding the orbital parameters of the binary system. 

 The lowest luminosity at which pulsations are detected in the light curve is estimated to be $\sim$7.3$\times$10$^{34}$ erg s$^{-1}$ (in 1-70 keV range, assuming a distance of 5.99 kpc). This finding is close to the luminosity value of 1.8$\times$10$^{34}$ erg s$^{-1}$ obtained from the spectral fitting in the 0.4-12 keV energy range for a distance of 4.5 kpc using 2017 XMM-Newton observation \citet{2018A&A...613A..52R}. This result signifies that the source has not reached the propeller regime state \citep{1975A&A....39..185I} at luminosity close to $\sim$7.3$\times$10$^{34}$ erg s$^{-1}$. Detection of pulsation at lower luminosity could be attributed to the long spin period of the neutron star ($\approx$374 s).  Using the values of magnetic field and spin period at a lower luminosity of the pulsar as 5.69$\times$10$^{12}$ G and 374.57 s, the limiting luminosity for the onset of the propeller regime \citep{2002ApJ...580..389C} is estimated to be 1.3$\times$10$^{33}$ erg s$^{-1}$. Previously it has been reported that sources with longer spin periods ($\geq$100 s) exhibit pulsations at lower luminosities (10$^{32-33}$ erg s$^{-1}$) because of their relatively higher co-rotation radius, although exceptions exist, such as GX~1+4 and OAO~1653-40 (\citealt{1983ApJ...270..711W,1986ApJ...308..669S} and references therein).

\begin{figure}
    \centering
    \hspace*{0.1cm}
    \includegraphics[trim={0 1.5cm 0 0.5cm},scale=1.2]{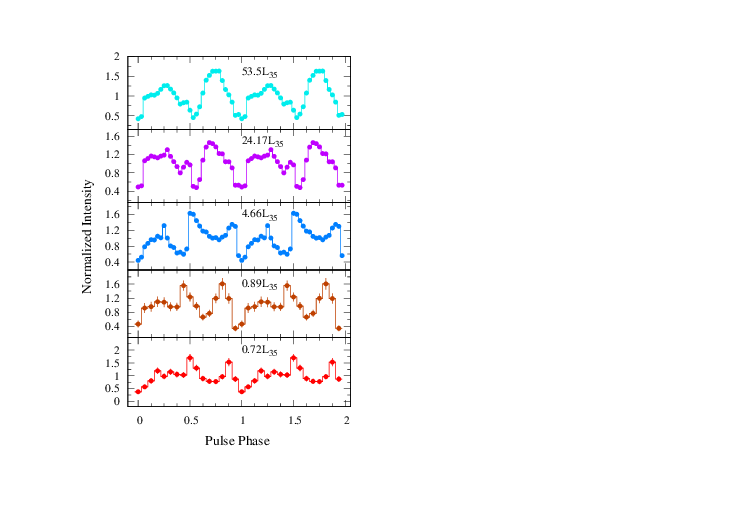}
    \caption{ Evolution of pulse profile in the 3-7 keV energy band with luminosity. }
    \label{fig:pp_lumin_var}
\end{figure}

Subsequently, we examined the pulse profiles of IGR~J06074+2205   obtained from NICER and NuSTAR observations to assess the modulation of X-ray photons due to the rotation of the neutron star. Pulse profiles contain information on the geometry of the X-ray emitting region and its surroundings. We analyze pulse profiles at different luminosities and energy ranges. The pulse profiles are found to be strongly dependent on luminosity and energy (Figure~\ref{fig:2idNICER_pp}, ~\ref{fig:niEreolved_pp}, \& ~\ref{fig:nuEreolved_pp}).  The pulse profiles obtained at higher luminosity exhibit smoother shapes compared to those obtained at lower luminosity, which possess a complex structure in the soft X-ray regime. For a clear understanding of the above fact, we chose an energy band that is common to both NICER and NuSTAR and has sufficient SNR. The luminosity evolution of the pulse profiles obtained in the 3-7 keV energy band are shown in Figure~\ref{fig:pp_lumin_var}. At lower luminosities, the pulse profiles are single-peak dominated and harbor deeper absorption features. However, as the luminosity increases, the pulse profiles evolve to a smoother double-peaked profile. We investigate the energy dependency of the pulse profiles in detail using NuSTAR data.  The pulse profile from the NuSTAR observation at a luminosity of approximately $5.56 \times 10^{36}$ erg s$^{-1}$ showed an asymmetric, smooth double-peaked structure in the 3-79 keV range, with the basic shape remaining consistent across various energy bands (Figure~\ref{fig:nuEreolved_pp}). As the energy increased from 3 to 40 keV, the peak in the 0.0-0.5 phase dimmed, while the peak in the 0.5-1.0 phase brightened. However, beyond 40 keV, the difference in the intensity of the two peaks started diminishing.  The pulse profile acquired at a luminosity of  4.66$\times$10$^{35}$ erg s$^{-1}$ also exhibited a double-peaked morphology, though accompanied by a complex second peak (Figure~\ref{fig:nuEreolved_pp}). For the lower luminosity case (red color profiles), a broad absorption dip is found in the second peak of the pulse profile below 20 keV. The 3depth of the dip is more pronounced in the 3-7 keV band and decreases with an increase in energy. Dips in the pulse profiles could be due to the absorption of low energy photons by accreting material locked in the magnetosphere asymmetrically \citep{1983ApJ...270..711W}. Dips in pulse profile due to absorption of radiation from the pulsar by matter stream in a specific phase of the magnetosphere is also observed in various other Be/X-ray binary pulsars such as V0332+53, 1A~0535+262, EXO~2030+375, GX~304-1, and RX~J0209.6-7427 \citep{2006MNRAS.371...19T, 2008ApJ...672..516N,2013ApJ...764..158N, 2015RAA....15..537N, Epili2017MNRAS.472.3455E, 2016MNRAS.457.2749J, 2020MNRAS.494.5350V,2024ApJ...963..132C}.  

The pulse profiles obtained from the NICER observations also exhibit distinct energy dependencies at different luminosity levels. The shape of the pulse profiles obtained at lower luminosity shows minimal variation with energy (see Figure~\ref{fig:2idNICER_pp}). Conversely, at a luminosity of  2.42$\times$10$^{36}$ erg s$^{-1}$, the pulse profiles are unique and show contrasting behavior relative to the lower luminosity case. The shape of the pulse profiles appears relatively smoother compared to the previously mentioned profiles, exhibiting a double-peaked morphology (Figure~\ref{fig:niEreolved_pp}). Notably, the pulse profile shows significant energy dependency. In the 0.5-1 keV band, the pulse profile consists of a single peak between 0.0-0.5 phase range (primary peak). As the energy increases, another peak arises in the 0.5-1 phase range (secondary peak). It is important to notice that the strength of the secondary peak keeps increasing with energy and eventually overtakes the primary peak at higher energy. The width of the secondary peak also keeps increasing and becomes comparable to the primary peak at higher energy. This type of energy dependency of pulse profiles is also reported by \citet{2024ATel16401....1N}. The evolution of pulse profiles in such a manner is observed in sources like GX~301-2 (\citealt{2004A&A...427..975K}) and Vela~X-1 (\citealt{2002A&A...395..129K}). However, this behavior of pulse profile is in contrast to many other accreting X-ray pulsars like GRO~J1008-57 \citep{2011MNRAS.413..241N}, 4U~0115+63 \citep{2007AstL...33..368T}, 4U~1909+07 \citep{2020MNRAS.498.4830J}, 4U~1901+03 \citep{2021Ap&SS.366...84R}, 1A~0535+262 \citep{2023MNRAS.518.5089C}, and 2S~1417-624 \citep{2018MNRAS.479.5612G}, where the secondary peak intensity decreases as the energy increases. By looking at the evolution of the pulse profile with energy, it can be inferred that the soft X-ray photons dominate the primary peak, whereas the secondary peak is due to the hard X-ray photons. The double-peaked profile might not be due to fan-beam emission as the source was emitting at a luminosity one order below the critical luminosity. This could be due to the contribution from two magnetic poles where soft X-ray absorbing materials have a contrasting distribution.

\begin{figure}
    \centering
    \includegraphics[trim={0 0.5cm 0 0.2cm},scale=0.6]{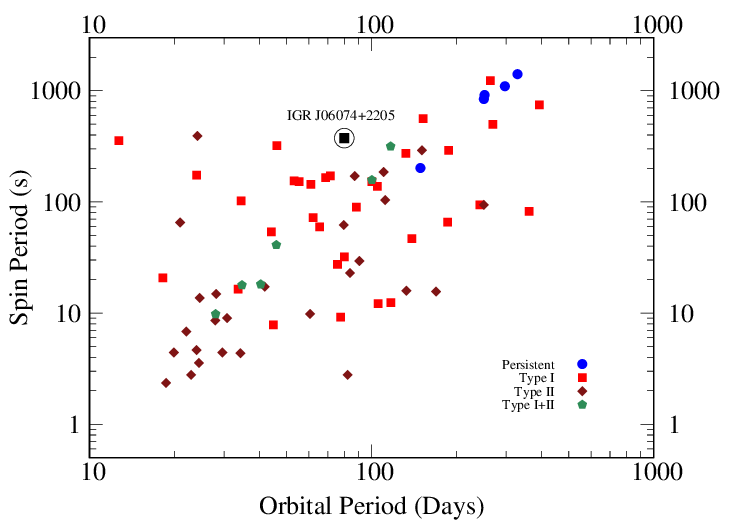}
    \caption{Diagram illustrating the orbital and spin periods of various BeXRBs (\citealt{2014ApJ...786..128C} and references therein). Different colors are employed to represent the outburst behavior observed for each source. The point representing the orbital period (80 days) and spin period (374.6 s) of IGR~J06074+2205 is shown in black color.}
    \label{fig:orbit_spin}
\end{figure}

 We also examine the variation of pulsed fraction (PF) with energy across two NuSTAR and three NICER observations (Figure~\ref{fig:pf_energy}). It is observed that the PF increases with energy \citep{1997ApJS..113..367B,2018MNRAS.474.4432J,2023MNRAS.521.3951J}. For ID~02, the PF initially rises steadily in the 3--30 keV range but drops sharply beyond 35 keV ~\citep{2024arXiv240217382R}. This behavior may be attributed to the presence of a cyclotron resonant scattering feature (CRSF), which is known to cause significant changes in pulse profiles, including pulse shape distortions, PF variations, and phase shifts near the cyclotron line energy \citep{2016MNRAS.457.2749J,2023A&A...677A.103F}. However, we cannot draw any definitive conclusions due to the lack of data beyond 45 keV. In contrast, the PF evolution follows a different trajectory for ID~04 when the source luminosity is low. This suggests complex changes in the emitting region with varying luminosity. The most intriguing case is ID~106, where a large hump-like structure is observed between 0.5 and 2.5 keV, followed by a steady increase in PF beyond 3 keV. This hump-like feature is absent in ID~105 and ID~111, suggesting that as luminosity increases, a pulsating component emerges in the soft X-ray regime.

We determine the possible orbital period of the source using the arguments detailed in Section~\ref{sub:orbital period}. The minimum difference between consecutive outbursts (Figure~\ref{fig:maxi_distribution}) is nearly 80 days. We calculated the difference (in days) between all consecutive outbursts and found that the differences are close to the multiples of 80 and 80/n (n=2,3,4) days. Hence, we infer that the possible orbital period of the binary is about 80 or 80/n (n=2,3,4) days. Similar results were also reported by \citet{2023ATel16351....1M} using data from four outbursts in 4-10 keV MAXI/GSC light curve. Considering the constraint on data quality and its availability at the current stage, this is the optimum we can infer on the orbital period of the binary. To contextualize this value within the Corbet diagram, we incorporated the spin and orbital period values into the plot assembled by \citet{2014ApJ...786..128C}, utilizing data from \citealt{2011MNRAS.416.1556T} (see Figure~\ref{fig:orbit_spin}).

We carried out spectral analysis during the October and December X-ray outbursts of the pulsar using NuSTAR and NICER observations.  During the NICER observations, the source luminosity varied between $\sim$5$\times$10$^{34}$-2.5$\times$10$^{36}$ erg s$^{-1}$ (for a distance of 5.99 kpc), with the spectra fitting well with an absorbed power-law model in 1-7 keV energy band. The temporal variation of the spectral parameters is illustrated in Figure~\ref{fig:NICER_spec}. The luminosity variation of the source over time indicates that NICER observed the source during the declining phases of both the outbursts. The photon index values varied between 1 and 2 without showing any particular trend. 

Accretion-powered binary X-ray pulsars possess a magnetic field of the order of 10$^{12}$ G. Hence, the CRSFs are expected to be detected in hard X-ray (10–100 keV) ranges \citep{2012A&A...544A.123B}. To investigate the presence of CRSF in IGR~J06074+2205, we carried out broadband spectral analysis using NuSTAR observations. All the continuum models describe the broadband spectra well, and the properties of the Gaussian absorption component in the spectrum representing the CRSF depend on the chosen continuum model. We choose the NPEX model as the best-fit model because the value of photon index (PI) and cutoff energy (E$_{cut}$) are constrained between 0.5–1.5 keV and 10–30 keV, respectively, after adding the Gaussian absorption component, which is typical for accreting pulsars \citep{1983ApJ...270..711W}. Considering the absorption component as the CRSF, the centroid energy of CRSF is found to be  $\sim$50.7 keV. We estimated the  magnetic field of the neutron star  using the relation  
\begin{equation}
    E_{CRSF}= 11.57 \times B_{12} (1+z)^{-1}
\end{equation}

where, $E_{CRSF}$ is the cyclotron line energy in keV, $B_{12}$ is the magnetic field in units of 10$^{12}$ G, and $z$ is the gravitational red-shift (z $\simeq$ 0.3 for typical neutron star). The calculated magnetic field of the pulsar is 5.69$\times$10$^{12}$ G. According to \citep{2012A&A...544A.123B}, the critical luminosity of an accretion-powered X-ray pulsar depends on the magnetic field and is given as

\begin{equation}
    L_{crit}= 1.49 \times 10^{37} (\frac{B}{10^{12} G})^{16/15}~ erg~s^{-1}
\end{equation}

Using Equation~4 and considering a neutron star of mass 1.4M$_{\odot}$, radius 10 km, and magnetic field  5.69$\times$10$^{12}$ G, the value of L$_{crit}$ is estimated to be around 9.52$\times 10^{37}$ erg s$^{-1}$. This signifies that the neutron star is accreting matter in the sub-critical regime.

To probe the dynamics of the circumstellar disc around the companion Be star in the binary system, we analyzed the optical spectra of  IGR~J06074+2205. We observed the H$\alpha$ line in emission, indicating the presence of circumstellar disc around the Be star \citep{2006MNRAS.368..447C,2010A&A...522A.107R,2023MNRAS.518.5089C,2024BSRSL..93..657N}. The equivalent width (EW) of the H$\alpha$ line and the disc radius are found to increase over time (Table~\ref{tab:optical parameters}). This signifies that a larger or denser disc is forming around the Be star with time. The H$\alpha$ line evolved from a double-peaked structure to a single-peak dominated structure between 18 November 2022 and 3 February 2024 (Figure~\ref{Fig:halpha_FeII_fitting}). The change in the $V/R$ value from 1.08 to 0.41 indicates the change in the structure of the line profile from blue-peak dominated to red-peak dominated shape. The FWHM of the  H$\alpha$ line varies within uncertainties throughout the observation duration and remains around $\approx$700 km s$^{-1}$.
 
We also observed a FeII line in emission at a wavelength of 5577.91 \AA. This is one of the rarest FeII emission lines observed at this wavelength in Be stars \citep{1992ApJS...81..335S,2011BASI...39..517M,2006A&A...450..427S}. The ionization potential of a neutral Fe atom is 7.8 eV, whereas for FeII ions, it is 16.2 eV. This indicates that FeII lines originate in regions close to the central star.  Interestingly, \citet{2010A&A...522A.107R} observed IGR~J06074+2205 between 2006 and 2010 and did not observe any signature of the FeII line. This indicates the transient nature of the FeII line.  The FeII line remained double-peaked throughout our monitoring period. The FeII emission lines observed on 22 October and 15 December 2023 exhibit emission in absorption (emission lines with underlying photospheric absorption; \citealt{2016RAA....16..138H,2022JApA...43..102B}) type profile. After 1.5 months, it evolved to a double-peaked emission type profile (Figure~\ref{Fig:halpha_FeII_fitting}). The double-peaked nature of FeII line also indicates its origin from the innermost region. The peak velocity difference between red-shifted and blue-shifted lines for FeII line varies within error bars throughout the observation duration and remains at around $\approx$380 km s$^{-1}$. The $V/R$ value changes from 1.53 to 0.71, suggesting changes in the profile structure from blue peak dominated to red peak dominated shape as in the case for the H$\alpha$ line. The size of the FeII emitting region is estimated to be around 4 R$_{*}$.
 
One interesting result we observed is the distinction in $V/R$ variability of H$\alpha$ and FeII lines. The $V/R$ variability suggests rotation of a one-armed perturbation (a zone in the disc with higher density than the rest of the disc) in the circumstellar disc \citep{1997A&A...318..548O, 1992A&A...265L..45P}. $V/R>$1 means the perturbed region is approaching the observer, and $V/R<$1 means the perturbed region is receding from the observer. We observed that the $V/R$ values for H$\alpha$ and FeII lines changed from greater than one to less than one on 13 December 2022 and 15 December 2023, respectively. The delay of about one year for two different emission lines indicates that the perturbed region is not symmetrically distributed in the radial direction. Different $V/R$ variabilities for various emission lines are also observed in 1A~0535+262 \citep{2013PASJ...65...83M}. The average $\Delta$V obtained for H$\alpha$ and FeII lines are 340 km s$^{-1}$ and 250 km s$^{-1}$, respectively. The average FWHM obtained for H$\alpha$ and FeII lines are $\sim$700 km s$^{-1}$ and $\sim$370 km s$^{-1}$, respectively. These results suggest that the particles emitting H$\alpha$ photons present in the higher velocity (disc region closer to central star) region \citep{1992ApJS...81..335S} compared to the FeII line emitting particles. The appearance of other emission lines like HeI (5875.72~\AA), HeI (6678~\AA), and HeI (7065~\AA) on the latter phases of our observation hinted towards the growing of larger or denser circumstellar disc around the Be star of the IGR~J06074+2205 binary.

 \citet{2024OEJV..249....1N} studied the long-term variability of IGR~J06074+2205 using various photometric filters covering the 4100-8100  \AA~ range. They identified long-term optical variability with a period of approximately 620 days, which they attributed to either the precession of the circumstellar disc or the propagation of a density wave within the disc. While comparing the timelines (see Figure~1 of \citealt{2024OEJV..249....1N}), we found that our first two optical observations were carried out during the decreasing phase of the V-band magnitude, while the subsequent three observations were made during the rising phase. Despite the V-band flux decreasing during the last three observations, the equivalent width (EW) of the H$\alpha$ line continued to increase over time. This behavior is consistent with findings by \citet{2024OEJV..249....1N}, who observed that optical variability and EW variability operate on different timescales. Furthermore, \citet{2010A&A...522A.107R} reported that the H$\alpha$ line in IGR~J06074+2205 exhibited a transition from emission to absorption, and during that period, no major X-ray outbursts occurred. This phenomenon could be explained by the precession of the circumstellar disc, as suggested by \citet{2024OEJV..249....1N}.

Furthermore, Figure~\ref{fig:optical_spectra} \& ~\ref{Fig:halpha_FeII_fitting} exhibit dynamics of the circumstellar disc around the Be star between November 2022 and February 2024. During the later part of our observation, we get the signatures of many emission lines in the optical spectra, although most of the lines are weak. However, if we look at the evolution of two strong emission lines, H$\alpha$ and FeII, we can notice that their EW increases with time. These facts highlight an essential point that the X-ray outburst that occurred during October and December 2023 (\citealt{2023ATel16351....1M},~\citealt{2023ATel16394....1N}, also see Figure~\ref{fig:log}) did not have any significant effect on the line emitting region of the circumstellar disc of the Be star as the disc is continuously becoming larger or denser. This behavior of the circumstellar disc is in contrast with that during the giant or Type~II X-ray outburst phase where a significant change in H$\alpha$ line EW and other properties are observed after the giant X-ray outbursts ~(e.g., 4U~0115+63; ~\citealt{2007A&A...462.1081R}, 1A~0535+262;~\citealt{2023MNRAS.518.5089C} ). The increase in the EW of the H$\alpha$ line suggests that the Be circumstellar disc is evolving continuously, which may lead to a giant outburst in the future in IGR~J06074+2205.

\section{Conclusion}
\label{conclusion}
 We carried out  X-ray studies of the Be/X-ray binary IGR~J06074+2205 during X-ray outbursts in October and December 2023 using NuSTAR and NICER observations of the pulsar. NuSTAR observed the source twice during the October 2023 outburst, while NICER provided coverage across various epochs during both outbursts. We observed coherent X-ray pulsations from the neutron star at $\sim$374.60 seconds. The pulse profiles of the pulsar exhibit a strong correlation with both luminosity and energy, revealing the intricate characteristics of the emitting region. In the low luminosity level, the pulse profiles are relatively complex compared to those at higher luminosity. The pulse profiles exhibit a dip in the soft X-ray band, which vanishes in the hard X-ray regime. Furthermore, the NuSTAR spectra unveil an iron emission line at around 6.4 keV during the brighter state, corresponding to a luminosity of approximately 5.56$\times$10$^{36}$ erg s$^{-1}$. Additionally, a cyclotron absorption line at $\sim$50.7 keV, indicative of magnetic field strength of 5.69$\times$10$^{12}$ G, is detected solely during this brighter observation. A simple absorbed power-law model adequately described NICER spectra within the 1-7 keV band. Expanding our analysis, we utilized the long-term MAXI/GSC light curve to estimate the potential orbital period of IGR~J06074+2205, which is predicted to be approximately 80 days or 80/n days (n=2,3,4). We showed results from optical spectroscopic analysis of observations taken between 2022 and 2024 using the MIRO and IAO. We observed variable H$\alpha$ and FeII emission lines, with an increase in equivalent width, indicating a dynamic circumstellar disc. Notable variations in the $V/R$ ratio for H$\alpha$ and FeII lines are also observed.  The appearance of additional emission lines, such as HeI (5875.72~\AA), HeI (6678~\AA), and HeI (7065~\AA)  from the post-outbursts observation in February 2024 suggests the growth of a larger or denser circumstellar disc. This disc continues to grow without noticeable mass loss, even during the 2023 X-ray outbursts, potentially leading to a future giant X-ray outburst.

\section*{Acknowledgements}
We thank the anonymous reviewer for the suggestions, which helped us to improve the manuscript.
The research work at the Physical Research Laboratory is funded by the Department of Space, Government of India. This research has made use of NuSTAR, NICER, and Fermi/GBM mission data and X-ray data analysis software provided by the High Energy Astrophysics Science Archive Research Center (HEASARC), which is a service of the Astrophysics Science Division at NASA/GSFC. This research has made use of MAXI data provided by RIKEN, JAXA and the MAXI team. We acknowledge the use of public data from the Swift data archive. The authors thank  MFOSC-P instrument team members of Physical Research Laboratory, India, for their constant support during the optical observations at various epochs. We also thank them for providing the data reduction pipeline and other support as and when required. We thank the staff of IAO, Hanle and CREST, Hosakote, who made these observations possible. The facilities at IAO and CREST are operated by the Indian Institute of Astrophysics, Bangalore. 

\section*{Data Availability}

We used the optical data from MIRO and IAO observatories in this work, respectively. The optical data can be shared on request. The X-ray data from NuSTAR\footnote{\url{https://heasarc.gsfc.nasa.gov/cgi-bin/W3Browse/w3browse.pl}}, NICER\footnote{\url{https://heasarc.gsfc.nasa.gov/cgi-bin/W3Browse/w3browse.pl}},  Fermi/GBM\footnote{\url{https://gammaray.nsstc.nasa.gov/gbm/science/pulsars/lightcurves/igrj06074.html}}, MAXI/GSC\footnote{\url{http://maxi.riken.jp/star_data/J0607+220/J0607+220.html}}, and Swift/BAT\footnote{\url{https://swift.gsfc.nasa.gov/results/transients/weak/IGRJ06074p2205}} data are publicly available.





\bibliographystyle{mnras}
\bibliography{IGRJ0} 

\begin{thebibliography}{}
\makeatletter
\relax
\def\mn@urlcharsother{\let\do\@makeother \do\$\do\&\do\#\do\^\do\_\do\%\do\~}
\def\mn@doi{\begingroup\mn@urlcharsother \@ifnextchar [ {\mn@doi@} {\mn@doi@[]}}
\def\mn@doi@[#1]#2{\def\@tempa{#1}\ifx\@tempa\@empty \href {http://dx.doi.org/#2} {doi:#2}\else \href {http://dx.doi.org/#2} {#1}\fi \endgroup}
\def\mn@eprint#1#2{\mn@eprint@#1:#2::\@nil}
\def\mn@eprint@arXiv#1{\href {http://arxiv.org/abs/#1} {{\tt arXiv:#1}}}
\def\mn@eprint@dblp#1{\href {http://dblp.uni-trier.de/rec/bibtex/#1.xml} {dblp:#1}}
\def\mn@eprint@#1:#2:#3:#4\@nil{\def\@tempa {#1}\def\@tempb {#2}\def\@tempc {#3}\ifx \@tempc \@empty \let \@tempc \@tempb \let \@tempb \@tempa \fi \ifx \@tempb \@empty \def\@tempb {arXiv}\fi \@ifundefined {mn@eprint@\@tempb}{\@tempb:\@tempc}{\expandafter \expandafter \csname mn@eprint@\@tempb\endcsname \expandafter{\@tempc}}}

\bibitem[\protect\citeauthoryear{{Angeloni} et~al.,}{{Angeloni} et~al.}{2019}]{2019AJ....157..156A}
{Angeloni} R.,  et~al., 2019, \mn@doi [\aj] {10.3847/1538-3881/ab0cf7}, \href {https://ui.adsabs.harvard.edu/abs/2019AJ....157..156A} {157, 156}

\bibitem[\protect\citeauthoryear{{Arias}, {Zorec}, {Cidale}, {Ringuelet}, {Morrell}  \& {Ballereau}}{{Arias} et~al.}{2006}]{2006A&A...460..821A}
{Arias} M.~L.,  {Zorec} J.,  {Cidale} L.,  {Ringuelet} A.~E.,  {Morrell} N.~I.,   {Ballereau} D.,  2006, \mn@doi [\aap] {10.1051/0004-6361:20065160}, \href {https://ui.adsabs.harvard.edu/abs/2006A&A...460..821A} {460, 821}

\bibitem[\protect\citeauthoryear{{Arnaud}}{{Arnaud}}{1996}]{1996ASPC..101...17A}
{Arnaud} K.~A.,  1996, in {Jacoby} G.~H.,  {Barnes} J.,  eds,  Astronomical Society of the Pacific Conference Series Vol. 101, Astronomical Data Analysis Software and Systems V. p.~17

\bibitem[\protect\citeauthoryear{{Banerjee} et~al.,}{{Banerjee} et~al.}{2022}]{2022JApA...43..102B}
{Banerjee} G.,  et~al., 2022, \mn@doi [Journal of Astrophysics and Astronomy] {10.1007/s12036-022-09891-y}, \href {https://ui.adsabs.harvard.edu/abs/2022JApA...43..102B} {43, 102}

\bibitem[\protect\citeauthoryear{{Becker} \& {Wolff}}{{Becker} \& {Wolff}}{2007}]{2007ApJ...654..435B}
{Becker} P.~A.,  {Wolff} M.~T.,  2007, \mn@doi [\apj] {10.1086/509108}, \href {https://ui.adsabs.harvard.edu/abs/2007ApJ...654..435B} {654, 435}

\bibitem[\protect\citeauthoryear{{Becker} et~al.,}{{Becker} et~al.}{2012}]{2012A&A...544A.123B}
{Becker} P.~A.,  et~al., 2012, \mn@doi [\aap] {10.1051/0004-6361/201219065}, \href {https://ui.adsabs.harvard.edu/abs/2012A&A...544A.123B} {544, A123}

\bibitem[\protect\citeauthoryear{{Bildsten} et~al.,}{{Bildsten} et~al.}{1997}]{1997ApJS..113..367B}
{Bildsten} L.,  et~al., 1997, \mn@doi [\apjs] {10.1086/313060}, \href {https://ui.adsabs.harvard.edu/abs/1997ApJS..113..367B} {113, 367}

\bibitem[\protect\citeauthoryear{{Campana}, {Stella}, {Israel}, {Moretti}, {Parmar}  \& {Orlandini}}{{Campana} et~al.}{2002}]{2002ApJ...580..389C}
{Campana} S.,  {Stella} L.,  {Israel} G.~L.,  {Moretti} A.,  {Parmar} A.~N.,   {Orlandini} M.,  2002, \mn@doi [\apj] {10.1086/343074}, \href {https://ui.adsabs.harvard.edu/abs/2002ApJ...580..389C} {580, 389}

\bibitem[\protect\citeauthoryear{{Chenevez}, {Budtz-Jorgensen}, {Lund}, {Westergaard}, {Kretschmar}, {Rodriguez}, {Orr}  \& {Hermsen}}{{Chenevez} et~al.}{2004}]{2004ATel..223....1C}
{Chenevez} J.,  {Budtz-Jorgensen} C.,  {Lund} N.,  {Westergaard} N.~J.,  {Kretschmar} P.,  {Rodriguez} J.,  {Orr} A.,   {Hermsen} W.,  2004, The Astronomer's Telegram, \href {https://ui.adsabs.harvard.edu/abs/2004ATel..223....1C} {223, 1}

\bibitem[\protect\citeauthoryear{{Cheng}, {Shao}  \& {Li}}{{Cheng} et~al.}{2014}]{2014ApJ...786..128C}
{Cheng} Z.~Q.,  {Shao} Y.,   {Li} X.~D.,  2014, \mn@doi [\apj] {10.1088/0004-637X/786/2/128}, \href {https://ui.adsabs.harvard.edu/abs/2014ApJ...786..128C} {786, 128}

\bibitem[\protect\citeauthoryear{{Chhotaray}, {Jaisawal}, {Kumari}, {Naik}, {Kumar}  \& {Jana}}{{Chhotaray} et~al.}{2023}]{2023MNRAS.518.5089C}
{Chhotaray} B.,  {Jaisawal} G.~K.,  {Kumari} N.,  {Naik} S.,  {Kumar} V.,   {Jana} A.,  2023, \mn@doi [\mnras] {10.1093/mnras/stac3354}, \href {https://ui.adsabs.harvard.edu/abs/2023MNRAS.518.5089C} {518, 5089}

\bibitem[\protect\citeauthoryear{{Chhotaray}, {Jaisawal}, {Nandi}, {Naik}, {Kumari}, {Ng}  \& {Gendreau}}{{Chhotaray} et~al.}{2024}]{2024ApJ...963..132C}
{Chhotaray} B.,  {Jaisawal} G.~K.,  {Nandi} P.,  {Naik} S.,  {Kumari} N.,  {Ng} M.,   {Gendreau} K.~C.,  2024, \mn@doi [\apj] {10.3847/1538-4357/ad235d}, \href {https://ui.adsabs.harvard.edu/abs/2024ApJ...963..132C} {963, 132}

\bibitem[\protect\citeauthoryear{{Coe}, {Reig}, {McBride}, {Galache}  \& {Fabregat}}{{Coe} et~al.}{2006}]{2006MNRAS.368..447C}
{Coe} M.~J.,  {Reig} P.,  {McBride} V.~A.,  {Galache} J.~L.,   {Fabregat} J.,  2006, \mn@doi [\mnras] {10.1111/j.1365-2966.2006.10127.x}, \href {https://ui.adsabs.harvard.edu/abs/2006MNRAS.368..447C} {368, 447}

\bibitem[\protect\citeauthoryear{{Corbet} et~al.,}{{Corbet} et~al.}{2022}]{2022ATel15614....1C}
{Corbet} R. H.~D.,  et~al., 2022, The Astronomer's Telegram, \href {https://ui.adsabs.harvard.edu/abs/2022ATel15614....1C} {15614, 1}

\bibitem[\protect\citeauthoryear{{Cowsik}, {Srinivasan}  \& {Prabhu}}{{Cowsik} et~al.}{2002}]{2002ASPC..266..424C}
{Cowsik} R.,  {Srinivasan} R.,   {Prabhu} T.,  2002, in {Vernin} J.,  {Benkhaldoun} Z.,   {Mu{\~n}oz-Tu{\~n}{\'o}n} C.,  eds,  Astronomical Society of the Pacific Conference Series Vol. 266, Astronomical Site Evaluation in the Visible and Radio Range. p.~424

\bibitem[\protect\citeauthoryear{{Dickey} \& {Lockman}}{{Dickey} \& {Lockman}}{1990}]{1990ARA&A..28..215D}
{Dickey} J.~M.,  {Lockman} F.~J.,  1990, \mn@doi [\araa] {10.1146/annurev.aa.28.090190.001243}, \href {https://ui.adsabs.harvard.edu/abs/1990ARA&A..28..215D} {28, 215}

\bibitem[\protect\citeauthoryear{{Epili}, {Naik}, {Jaisawal}  \& {Gupta}}{{Epili} et~al.}{2017}]{Epili2017MNRAS.472.3455E}
{Epili} P.,  {Naik} S.,  {Jaisawal} G.~K.,   {Gupta} S.,  2017, \mn@doi [\mnras] {10.1093/mnras/stx2247}, \href {https://ui.adsabs.harvard.edu/abs/2017MNRAS.472.3455E} {472, 3455}

\bibitem[\protect\citeauthoryear{{Ferrigno}, {D'A{\`\i}}  \& {Ambrosi}}{{Ferrigno} et~al.}{2023}]{2023A&A...677A.103F}
{Ferrigno} C.,  {D'A{\`\i}} A.,   {Ambrosi} E.,  2023, \mn@doi [\aap] {10.1051/0004-6361/202347062}, \href {https://ui.adsabs.harvard.edu/abs/2023A&A...677A.103F} {677, A103}

\bibitem[\protect\citeauthoryear{{Fortin}, {Garc{\'\i}a}  \& {Chaty}}{{Fortin} et~al.}{2022}]{2022A&A...665A..69F}
{Fortin} F.,  {Garc{\'\i}a} F.,   {Chaty} S.,  2022, \mn@doi [\aap] {10.1051/0004-6361/202244048}, \href {https://ui.adsabs.harvard.edu/abs/2022A&A...665A..69F} {665, A69}

\bibitem[\protect\citeauthoryear{{Gendreau} et~al.,}{{Gendreau} et~al.}{2016}]{2016SPIE.9905E..1HG}
{Gendreau} K.~C.,  et~al., 2016, in {den Herder} J.-W.~A.,  {Takahashi} T.,   {Bautz} M.,  eds,  Society of Photo-Optical Instrumentation Engineers (SPIE) Conference Series Vol. 9905, Space Telescopes and Instrumentation 2016: Ultraviolet to Gamma Ray. p. 99051H, \mn@doi{10.1117/12.2231304}

\bibitem[\protect\citeauthoryear{{Gupta}, {Naik}, {Jaisawal}  \& {Epili}}{{Gupta} et~al.}{2018}]{2018MNRAS.479.5612G}
{Gupta} S.,  {Naik} S.,  {Jaisawal} G.~K.,   {Epili} P.~R.,  2018, \mn@doi [\mnras] {10.1093/mnras/sty1804}, \href {https://ui.adsabs.harvard.edu/abs/2018MNRAS.479.5612G} {479, 5612}

\bibitem[\protect\citeauthoryear{{HI4PI Collaboration} et~al.,}{{HI4PI Collaboration} et~al.}{2016}]{2016A&A...594A.116H}
{HI4PI Collaboration} et~al., 2016, \mn@doi [\aap] {10.1051/0004-6361/201629178}, \href {https://ui.adsabs.harvard.edu/abs/2016A&A...594A.116H} {594, A116}

\bibitem[\protect\citeauthoryear{{Halpern} \& {Tyagi}}{{Halpern} \& {Tyagi}}{2005}]{2005ATel..682....1H}
{Halpern} J.~P.,  {Tyagi} S.,  2005, The Astronomer's Telegram, \href {https://ui.adsabs.harvard.edu/abs/2005ATel..682....1H} {682, 1}

\bibitem[\protect\citeauthoryear{{Harrison} et~al.,}{{Harrison} et~al.}{2013}]{2013ApJ...770..103H}
{Harrison} F.~A.,  et~al., 2013, \mn@doi [\apj] {10.1088/0004-637X/770/2/103}, \href {https://ui.adsabs.harvard.edu/abs/2013ApJ...770..103H} {770, 103}

\bibitem[\protect\citeauthoryear{{Herbig}}{{Herbig}}{1995}]{1995ARA&A..33...19H}
{Herbig} G.~H.,  1995, \mn@doi [\araa] {10.1146/annurev.aa.33.090195.000315}, \href {https://ui.adsabs.harvard.edu/abs/1995ARA&A..33...19H} {33, 19}

\bibitem[\protect\citeauthoryear{{Hou} et~al.,}{{Hou} et~al.}{2016}]{2016RAA....16..138H}
{Hou} W.,  et~al., 2016, \mn@doi [Research in Astronomy and Astrophysics] {10.1088/1674-4527/16/9/138}, \href {https://ui.adsabs.harvard.edu/abs/2016RAA....16..138H} {16, 138}

\bibitem[\protect\citeauthoryear{{Huang}}{{Huang}}{1972}]{1972ApJ...171..549H}
{Huang} S.-S.,  1972, \mn@doi [\apj] {10.1086/151309}, \href {https://ui.adsabs.harvard.edu/abs/1972ApJ...171..549H} {171, 549}

\bibitem[\protect\citeauthoryear{{Illarionov} \& {Sunyaev}}{{Illarionov} \& {Sunyaev}}{1975}]{1975A&A....39..185I}
{Illarionov} A.~F.,  {Sunyaev} R.~A.,  1975, \aap, \href {https://ui.adsabs.harvard.edu/abs/1975A&A....39..185I} {39, 185}

\bibitem[\protect\citeauthoryear{{Jaisawal}, {Naik}  \& {Epili}}{{Jaisawal} et~al.}{2016}]{2016MNRAS.457.2749J}
{Jaisawal} G.~K.,  {Naik} S.,   {Epili} P.,  2016, \mn@doi [\mnras] {10.1093/mnras/stw085}, \href {https://ui.adsabs.harvard.edu/abs/2016MNRAS.457.2749J} {457, 2749}

\bibitem[\protect\citeauthoryear{{Jaisawal}, {Naik}  \& {Chenevez}}{{Jaisawal} et~al.}{2018}]{2018MNRAS.474.4432J}
{Jaisawal} G.~K.,  {Naik} S.,   {Chenevez} J.,  2018, \mn@doi [\mnras] {10.1093/mnras/stx3082}, \href {https://ui.adsabs.harvard.edu/abs/2018MNRAS.474.4432J} {474, 4432}

\bibitem[\protect\citeauthoryear{{Jaisawal}, {Naik}, {Ho}, {Kumari}, {Epili}  \& {Vasilopoulos}}{{Jaisawal} et~al.}{2020}]{2020MNRAS.498.4830J}
{Jaisawal} G.~K.,  {Naik} S.,  {Ho} W. C.~G.,  {Kumari} N.,  {Epili} P.,   {Vasilopoulos} G.,  2020, \mn@doi [\mnras] {10.1093/mnras/staa2604}, \href {https://ui.adsabs.harvard.edu/abs/2020MNRAS.498.4830J} {498, 4830}

\bibitem[\protect\citeauthoryear{{Jaisawal} et~al.,}{{Jaisawal} et~al.}{2023}]{2023MNRAS.521.3951J}
{Jaisawal} G.~K.,  et~al., 2023, \mn@doi [\mnras] {10.1093/mnras/stad781}, \href {https://ui.adsabs.harvard.edu/abs/2023MNRAS.521.3951J} {521, 3951}

\bibitem[\protect\citeauthoryear{{Kalberla}, {Burton}, {Hartmann}, {Arnal}, {Bajaja}, {Morras}  \& {P{\"o}ppel}}{{Kalberla} et~al.}{2005}]{2005A&A...440..775K}
{Kalberla} P.~M.~W.,  {Burton} W.~B.,  {Hartmann} D.,  {Arnal} E.~M.,  {Bajaja} E.,  {Morras} R.,   {P{\"o}ppel} W.~G.~L.,  2005, \mn@doi [\aap] {10.1051/0004-6361:20041864}, \href {https://ui.adsabs.harvard.edu/abs/2005A&A...440..775K} {440, 775}

\bibitem[\protect\citeauthoryear{{Kre{\L}owski} \& {Schmidt}}{{Kre{\L}owski} \& {Schmidt}}{1997}]{1997ApJ...477..209K}
{Kre{\L}owski} J.,  {Schmidt} M.,  1997, \mn@doi [\apj] {10.1086/303695}, \href {https://ui.adsabs.harvard.edu/abs/1997ApJ...477..209K} {477, 209}

\bibitem[\protect\citeauthoryear{{Kreykenbohm}, {Coburn}, {Wilms}, {Kretschmar}, {Staubert}, {Heindl}  \& {Rothschild}}{{Kreykenbohm} et~al.}{2002}]{2002A&A...395..129K}
{Kreykenbohm} I.,  {Coburn} W.,  {Wilms} J.,  {Kretschmar} P.,  {Staubert} R.,  {Heindl} W.~A.,   {Rothschild} R.~E.,  2002, \mn@doi [\aap] {10.1051/0004-6361:20021181}, \href {https://ui.adsabs.harvard.edu/abs/2002A&A...395..129K} {395, 129}

\bibitem[\protect\citeauthoryear{{Kreykenbohm}, {Wilms}, {Coburn}, {Kuster}, {Rothschild}, {Heindl}, {Kretschmar}  \& {Staubert}}{{Kreykenbohm} et~al.}{2004}]{2004A&A...427..975K}
{Kreykenbohm} I.,  {Wilms} J.,  {Coburn} W.,  {Kuster} M.,  {Rothschild} R.~E.,  {Heindl} W.~A.,  {Kretschmar} P.,   {Staubert} R.,  2004, \mn@doi [\aap] {10.1051/0004-6361:20035836}, \href {https://ui.adsabs.harvard.edu/abs/2004A&A...427..975K} {427, 975}

\bibitem[\protect\citeauthoryear{{Kumar} et~al.,}{{Kumar} et~al.}{2022}]{2022MNRAS.510.4265K}
{Kumar} V.,  et~al., 2022, \mn@doi [\mnras] {10.1093/mnras/stab3772}, \href {https://ui.adsabs.harvard.edu/abs/2022MNRAS.510.4265K} {510, 4265}

\bibitem[\protect\citeauthoryear{{Leahy}}{{Leahy}}{1987}]{1987A&A...180..275L}
{Leahy} D.~A.,  1987, \aap, \href {https://ui.adsabs.harvard.edu/abs/1987A&A...180..275L} {180, 275}

\bibitem[\protect\citeauthoryear{{Makishima}, {Mihara}, {Nagase}  \& {Tanaka}}{{Makishima} et~al.}{1999}]{1999ApJ...525..978M}
{Makishima} K.,  {Mihara} T.,  {Nagase} F.,   {Tanaka} Y.,  1999, \mn@doi [\apj] {10.1086/307912}, \href {https://ui.adsabs.harvard.edu/abs/1999ApJ...525..978M} {525, 978}

\bibitem[\protect\citeauthoryear{{Malacaria}, {Jenke}, {Roberts}, {Wilson-Hodge}, {Cleveland}, {Mailyan}  \& {GBM Accreting Pulsars Program Team}}{{Malacaria} et~al.}{2020}]{2020ApJ...896...90M}
{Malacaria} C.,  {Jenke} P.,  {Roberts} O.~J.,  {Wilson-Hodge} C.~A.,  {Cleveland} W.~H.,  {Mailyan} B.,   {GBM Accreting Pulsars Program Team} 2020, \mn@doi [\apj] {10.3847/1538-4357/ab855c}, \href {https://ui.adsabs.harvard.edu/abs/2020ApJ...896...90M} {896, 90}

\bibitem[\protect\citeauthoryear{{Mathew} \& {Subramaniam}}{{Mathew} \& {Subramaniam}}{2011}]{2011BASI...39..517M}
{Mathew} B.,  {Subramaniam} A.,  2011, \mn@doi [Bulletin of the Astronomical Society of India] {10.48550/arXiv.1108.5850}, \href {https://ui.adsabs.harvard.edu/abs/2011BASI...39..517M} {39, 517}

\bibitem[\protect\citeauthoryear{{Mihara} et~al.,}{{Mihara} et~al.}{2023}]{2023ATel16351....1M}
{Mihara} T.,  et~al., 2023, The Astronomer's Telegram, \href {https://ui.adsabs.harvard.edu/abs/2023ATel16351....1M} {16351, 1}

\bibitem[\protect\citeauthoryear{{Moritani} et~al.,}{{Moritani} et~al.}{2013}]{2013PASJ...65...83M}
{Moritani} Y.,  et~al., 2013, \mn@doi [\pasj] {10.1093/pasj/65.4.83}, \href {https://ui.adsabs.harvard.edu/abs/2013PASJ...65...83M} {65, 83}

\bibitem[\protect\citeauthoryear{{Naik} \& {Jaisawal}}{{Naik} \& {Jaisawal}}{2015}]{2015RAA....15..537N}
{Naik} S.,  {Jaisawal} G.~K.,  2015, \mn@doi [Research in Astronomy and Astrophysics] {10.1088/1674-4527/15/4/007}, \href {https://ui.adsabs.harvard.edu/abs/2015RAA....15..537N} {15, 537}

\bibitem[\protect\citeauthoryear{{Naik} et~al.,}{{Naik} et~al.}{2008}]{2008ApJ...672..516N}
{Naik} S.,  et~al., 2008, \mn@doi [\apj] {10.1086/523295}, \href {https://ui.adsabs.harvard.edu/abs/2008ApJ...672..516N} {672, 516}

\bibitem[\protect\citeauthoryear{{Naik}, {Paul}, {Kachhara}  \& {Vadawale}}{{Naik} et~al.}{2011}]{2011MNRAS.413..241N}
{Naik} S.,  {Paul} B.,  {Kachhara} C.,   {Vadawale} S.~V.,  2011, \mn@doi [\mnras] {10.1111/j.1365-2966.2010.18128.x}, \href {https://ui.adsabs.harvard.edu/abs/2011MNRAS.413..241N} {413, 241}

\bibitem[\protect\citeauthoryear{{Naik}, {Maitra}, {Jaisawal}  \& {Paul}}{{Naik} et~al.}{2013}]{2013ApJ...764..158N}
{Naik} S.,  {Maitra} C.,  {Jaisawal} G.~K.,   {Paul} B.,  2013, \mn@doi [\apj] {10.1088/0004-637X/764/2/158}, \href {https://ui.adsabs.harvard.edu/abs/2013ApJ...764..158N} {764, 158}

\bibitem[\protect\citeauthoryear{{Naik}, {Chhotaray}  \& {Kumari}}{{Naik} et~al.}{2024}]{2024BSRSL..93..657N}
{Naik} S.,  {Chhotaray} B.,   {Kumari} N.,  2024, \mn@doi [Bulletin de la Societe Royale des Sciences de Liege] {10.25518/0037-9565.11829}, \href {https://ui.adsabs.harvard.edu/abs/2024BSRSL..93..657N} {93, 657}

\bibitem[\protect\citeauthoryear{{Nakajima} et~al.,}{{Nakajima} et~al.}{2023}]{2023ATel16394....1N}
{Nakajima} M.,  et~al., 2023, The Astronomer's Telegram, \href {https://ui.adsabs.harvard.edu/abs/2023ATel16394....1N} {16394, 1}

\bibitem[\protect\citeauthoryear{{Nakajima}, {Negoro}, {Mihara}, {Sugizaki}, {Iwakiri}, {Gendreau}  \& {Arzoumanian}}{{Nakajima} et~al.}{2024}]{2024ATel16401....1N}
{Nakajima} M.,  {Negoro} H.,  {Mihara} T.,  {Sugizaki} M.,  {Iwakiri} W.,  {Gendreau} K.,   {Arzoumanian} Z.,  2024, The Astronomer's Telegram, \href {https://ui.adsabs.harvard.edu/abs/2024ATel16401....1N} {16401, 1}

\bibitem[\protect\citeauthoryear{{Negueruela}}{{Negueruela}}{1998}]{1998A&A...338..505N}
{Negueruela} I.,  1998, \mn@doi [\aap] {10.48550/arXiv.astro-ph/9807158}, \href {https://ui.adsabs.harvard.edu/abs/1998A&A...338..505N} {338, 505}

\bibitem[\protect\citeauthoryear{{Nesci}, {Fiocchi}  \& {Vagnozzi}}{{Nesci} et~al.}{2024}]{2024OEJV..249....1N}
{Nesci} R.,  {Fiocchi} M.,   {Vagnozzi} A.,  2024, \mn@doi [Open European Journal on Variable Stars] {10.5817/OEJV2024-0249}, \href {https://ui.adsabs.harvard.edu/abs/2024OEJV..249....1N} {249, 1}

\bibitem[\protect\citeauthoryear{{Okazaki}}{{Okazaki}}{1997}]{1997A&A...318..548O}
{Okazaki} A.~T.,  1997, \aap, \href {https://ui.adsabs.harvard.edu/abs/1997A&A...318..548O} {318, 548}

\bibitem[\protect\citeauthoryear{{Papaloizou}, {Savonije}  \& {Henrichs}}{{Papaloizou} et~al.}{1992}]{1992A&A...265L..45P}
{Papaloizou} J.~C.,  {Savonije} G.~J.,   {Henrichs} H.~F.,  1992, \aap, \href {https://ui.adsabs.harvard.edu/abs/1992A&A...265L..45P} {265, L45}

\bibitem[\protect\citeauthoryear{{Paul} \& {Naik}}{{Paul} \& {Naik}}{2011}]{2011BASI...39..429P}
{Paul} B.,  {Naik} S.,  2011, \mn@doi [Bulletin of the Astronomical Society of India] {10.48550/arXiv.1110.4446}, \href {https://ui.adsabs.harvard.edu/abs/2011BASI...39..429P} {39, 429}

\bibitem[\protect\citeauthoryear{{Pooley}}{{Pooley}}{2004}]{2004ATel..226....1P}
{Pooley} G.,  2004, The Astronomer's Telegram, \href {https://ui.adsabs.harvard.edu/abs/2004ATel..226....1P} {226, 1}

\bibitem[\protect\citeauthoryear{{Porter} \& {Rivinius}}{{Porter} \& {Rivinius}}{2003}]{2003PASP..115.1153P}
{Porter} J.~M.,  {Rivinius} T.,  2003, \mn@doi [\pasp] {10.1086/378307}, \href {https://ui.adsabs.harvard.edu/abs/2003PASP..115.1153P} {115, 1153}

\bibitem[\protect\citeauthoryear{{Prigozhin} et~al.,}{{Prigozhin} et~al.}{2016}]{2016SPIE.9905E..1IP}
{Prigozhin} G.,  et~al., 2016, in {den Herder} J.-W.~A.,  {Takahashi} T.,   {Bautz} M.,  eds,  Society of Photo-Optical Instrumentation Engineers (SPIE) Conference Series Vol. 9905, Space Telescopes and Instrumentation 2016: Ultraviolet to Gamma Ray. p. 99051I, \mn@doi{10.1117/12.2231718}

\bibitem[\protect\citeauthoryear{{Rai} \& {Paul}}{{Rai} \& {Paul}}{2021}]{2021Ap&SS.366...84R}
{Rai} B.,  {Paul} B.~C.,  2021, \mn@doi [\apss] {10.1007/s10509-021-03971-1}, \href {https://ui.adsabs.harvard.edu/abs/2021Ap&SS.366...84R} {366, 84}

\bibitem[\protect\citeauthoryear{{Reig}}{{Reig}}{2011}]{2011Ap&SS.332....1R}
{Reig} P.,  2011, \mn@doi [\apss] {10.1007/s10509-010-0575-8}, \href {https://ui.adsabs.harvard.edu/abs/2011Ap&SS.332....1R} {332, 1}

\bibitem[\protect\citeauthoryear{{Reig} \& {Zezas}}{{Reig} \& {Zezas}}{2018}]{2018A&A...613A..52R}
{Reig} P.,  {Zezas} A.,  2018, \mn@doi [\aap] {10.1051/0004-6361/201732533}, \href {https://ui.adsabs.harvard.edu/abs/2018A&A...613A..52R} {613, A52}

\bibitem[\protect\citeauthoryear{{Reig}, {Larionov}, {Negueruela}, {Arkharov}  \& {Kudryavtseva}}{{Reig} et~al.}{2007}]{2007A&A...462.1081R}
{Reig} P.,  {Larionov} V.,  {Negueruela} I.,  {Arkharov} A.~A.,   {Kudryavtseva} N.~A.,  2007, \mn@doi [\aap] {10.1051/0004-6361:20066217}, \href {https://ui.adsabs.harvard.edu/abs/2007A&A...462.1081R} {462, 1081}

\bibitem[\protect\citeauthoryear{{Reig}, {Zezas}  \& {Gkouvelis}}{{Reig} et~al.}{2010}]{2010A&A...522A.107R}
{Reig} P.,  {Zezas} A.,   {Gkouvelis} L.,  2010, \mn@doi [\aap] {10.1051/0004-6361/201014788}, \href {https://ui.adsabs.harvard.edu/abs/2010A&A...522A.107R} {522, A107}

\bibitem[\protect\citeauthoryear{{Remillard} et~al.,}{{Remillard} et~al.}{2022}]{2022AJ....163..130R}
{Remillard} R.~A.,  et~al., 2022, \mn@doi [\aj] {10.3847/1538-3881/ac4ae6}, \href {https://ui.adsabs.harvard.edu/abs/2022AJ....163..130R} {163, 130}

\bibitem[\protect\citeauthoryear{{Roy}, {Sharma}, {Manikantan}  \& {Paul}}{{Roy} et~al.}{2024}]{2024arXiv240217382R}
{Roy} K.,  {Sharma} R.,  {Manikantan} H.,   {Paul} B.,  2024, \mn@doi [arXiv e-prints] {10.48550/arXiv.2402.17382}, \href {https://ui.adsabs.harvard.edu/abs/2024arXiv240217382R} {p. arXiv:2402.17382}

\bibitem[\protect\citeauthoryear{{Saad} et~al.,}{{Saad} et~al.}{2006}]{2006A&A...450..427S}
{Saad} S.~M.,  et~al., 2006, \mn@doi [\aap] {10.1051/0004-6361:20041877}, \href {https://ui.adsabs.harvard.edu/abs/2006A&A...450..427S} {450, 427}

\bibitem[\protect\citeauthoryear{{Schmidtke}, {Hunter}  \& {Cowley}}{{Schmidtke} et~al.}{2024}]{2024ATel16560....1S}
{Schmidtke} P.~C.,  {Hunter} T.~B.,   {Cowley} A.~P.,  2024, The Astronomer's Telegram, \href {https://ui.adsabs.harvard.edu/abs/2024ATel16560....1S} {16560, 1}

\bibitem[\protect\citeauthoryear{{Simon}, {Bondar}  \& {Metlova}}{{Simon} et~al.}{2012}]{2012ATel.4172....1S}
{Simon} A.,  {Bondar} A.~V.,   {Metlova} N.~V.,  2012, The Astronomer's Telegram, \href {https://ui.adsabs.harvard.edu/abs/2012ATel.4172....1S} {4172, 1}

\bibitem[\protect\citeauthoryear{{Slettebak}, {Collins}  \& {Truax}}{{Slettebak} et~al.}{1992}]{1992ApJS...81..335S}
{Slettebak} A.,  {Collins} George~W. I.,   {Truax} R.,  1992, \mn@doi [\apjs] {10.1086/191696}, \href {https://ui.adsabs.harvard.edu/abs/1992ApJS...81..335S} {81, 335}

\bibitem[\protect\citeauthoryear{{Srivastava}, {Jangra}, {Dixit}, {Munjal}, {Arora}  \& {Mavani}}{{Srivastava} et~al.}{2018}]{2018SPIE10702E..4IS}
{Srivastava} M.~K.,  {Jangra} M.,  {Dixit} V.,  {Munjal} B.~S.,  {Arora} H.,   {Mavani} T.,  2018, in {Evans} C.~J.,  {Simard} L.,   {Takami} H.,  eds,  Society of Photo-Optical Instrumentation Engineers (SPIE) Conference Series Vol. 10702, Ground-based and Airborne Instrumentation for Astronomy VII. p. 107024I, \mn@doi{10.1117/12.2309306}

\bibitem[\protect\citeauthoryear{Srivastava, Kumar, Dixit, Patel, Jangra, Rajpurohit  \& Mathur}{Srivastava et~al.}{2021}]{srivastava2021design}
Srivastava M.~K.,  Kumar V.,  Dixit V.,  Patel A.,  Jangra M.,  Rajpurohit A.,   Mathur S.,  2021, Experimental Astronomy, 51, 345

\bibitem[\protect\citeauthoryear{{Staubert} et~al.,}{{Staubert} et~al.}{2019}]{2019A&A...622A..61S}
{Staubert} R.,  et~al., 2019, \mn@doi [\aap] {10.1051/0004-6361/201834479}, \href {https://ui.adsabs.harvard.edu/abs/2019A&A...622A..61S} {622, A61}

\bibitem[\protect\citeauthoryear{{Stella}, {White}  \& {Rosner}}{{Stella} et~al.}{1986}]{1986ApJ...308..669S}
{Stella} L.,  {White} N.~E.,   {Rosner} R.,  1986, \mn@doi [\apj] {10.1086/164538}, \href {https://ui.adsabs.harvard.edu/abs/1986ApJ...308..669S} {308, 669}

\bibitem[\protect\citeauthoryear{{Stellingwerf}}{{Stellingwerf}}{1978}]{1978ApJ...224..953S}
{Stellingwerf} R.~F.,  1978, \mn@doi [\apj] {10.1086/156444}, \href {https://ui.adsabs.harvard.edu/abs/1978ApJ...224..953S} {224, 953}

\bibitem[\protect\citeauthoryear{{Tobrej}, {Rai}, {Ghising}  \& {Paul}}{{Tobrej} et~al.}{2024}]{2024JHEAp..42..129T}
{Tobrej} M.,  {Rai} B.,  {Ghising} M.,   {Paul} B.~C.,  2024, \mn@doi [Journal of High Energy Astrophysics] {10.1016/j.jheap.2024.04.006}, \href {https://ui.adsabs.harvard.edu/abs/2024JHEAp..42..129T} {42, 129}

\bibitem[\protect\citeauthoryear{{Tomsick}, {Chaty}, {Rodriguez}, {Walter}  \& {Kaaret}}{{Tomsick} et~al.}{2006}]{2006ATel..959....1T}
{Tomsick} J.~A.,  {Chaty} S.,  {Rodriguez} J.,  {Walter} R.,   {Kaaret} P.,  2006, The Astronomer's Telegram, \href {https://ui.adsabs.harvard.edu/abs/2006ATel..959....1T} {959, 1}

\bibitem[\protect\citeauthoryear{{Townsend}, {Coe}, {Corbet}  \& {Hill}}{{Townsend} et~al.}{2011}]{2011MNRAS.416.1556T}
{Townsend} L.~J.,  {Coe} M.~J.,  {Corbet} R.~H.~D.,   {Hill} A.~B.,  2011, \mn@doi [\mnras] {10.1111/j.1365-2966.2011.19153.x}, \href {https://ui.adsabs.harvard.edu/abs/2011MNRAS.416.1556T} {416, 1556}

\bibitem[\protect\citeauthoryear{{Truemper}, {Pietsch}, {Reppin}, {Voges}, {Staubert}  \& {Kendziorra}}{{Truemper} et~al.}{1978}]{1978ApJ...219L.105T}
{Truemper} J.,  {Pietsch} W.,  {Reppin} C.,  {Voges} W.,  {Staubert} R.,   {Kendziorra} E.,  1978, \mn@doi [\apjl] {10.1086/182617}, \href {https://ui.adsabs.harvard.edu/abs/1978ApJ...219L.105T} {219, L105}

\bibitem[\protect\citeauthoryear{{Tsygankov}, {Lutovinov}, {Churazov}  \& {Sunyaev}}{{Tsygankov} et~al.}{2006}]{2006MNRAS.371...19T}
{Tsygankov} S.~S.,  {Lutovinov} A.~A.,  {Churazov} E.~M.,   {Sunyaev} R.~A.,  2006, \mn@doi [\mnras] {10.1111/j.1365-2966.2006.10610.x}, \href {https://ui.adsabs.harvard.edu/abs/2006MNRAS.371...19T} {371, 19}

\bibitem[\protect\citeauthoryear{{Tsygankov}, {Lutovinov}, {Churazov}  \& {Sunyaev}}{{Tsygankov} et~al.}{2007}]{2007AstL...33..368T}
{Tsygankov} S.~S.,  {Lutovinov} A.~A.,  {Churazov} E.~M.,   {Sunyaev} R.~A.,  2007, \mn@doi [Astronomy Letters] {10.1134/S1063773707060023}, \href {https://ui.adsabs.harvard.edu/abs/2007AstL...33..368T} {33, 368}

\bibitem[\protect\citeauthoryear{{Vasilopoulos} et~al.,}{{Vasilopoulos} et~al.}{2020}]{2020MNRAS.494.5350V}
{Vasilopoulos} G.,  et~al., 2020, \mn@doi [\mnras] {10.1093/mnras/staa991}, \href {https://ui.adsabs.harvard.edu/abs/2020MNRAS.494.5350V} {494, 5350}

\bibitem[\protect\citeauthoryear{{Verner}, {Ferland}, {Korista}  \& {Yakovlev}}{{Verner} et~al.}{1996}]{1996ApJ...465..487V}
{Verner} D.~A.,  {Ferland} G.~J.,  {Korista} K.~T.,   {Yakovlev} D.~G.,  1996, \mn@doi [\apj] {10.1086/177435}, \href {https://ui.adsabs.harvard.edu/abs/1996ApJ...465..487V} {465, 487}

\bibitem[\protect\citeauthoryear{{White}, {Swank}  \& {Holt}}{{White} et~al.}{1983}]{1983ApJ...270..711W}
{White} N.~E.,  {Swank} J.~H.,   {Holt} S.~S.,  1983, \mn@doi [\apj] {10.1086/161162}, \href {https://ui.adsabs.harvard.edu/abs/1983ApJ...270..711W} {270, 711}

\bibitem[\protect\citeauthoryear{{Wilms}, {Allen}  \& {McCray}}{{Wilms} et~al.}{2000}]{2000ApJ...542..914W}
{Wilms} J.,  {Allen} A.,   {McCray} R.,  2000, \mn@doi [\apj] {10.1086/317016}, \href {https://ui.adsabs.harvard.edu/abs/2000ApJ...542..914W} {542, 914}

\bibitem[\protect\citeauthoryear{{Zamanov}, {Stoyanov}, {Mart{\'\i}}, {Tomov}, {Belcheva}, {Luque-Escamilla}  \& {Latev}}{{Zamanov} et~al.}{2013}]{2013A&A...559A..87Z}
{Zamanov} R.,  {Stoyanov} K.,  {Mart{\'\i}} J.,  {Tomov} N.~A.,  {Belcheva} G.,  {Luque-Escamilla} P.~L.,   {Latev} G.,  2013, \mn@doi [\aap] {10.1051/0004-6361/201321991}, \href {https://ui.adsabs.harvard.edu/abs/2013A&A...559A..87Z} {559, A87}

\bibitem[\protect\citeauthoryear{{Zechmeister} \& {K{\"u}rster}}{{Zechmeister} \& {K{\"u}rster}}{2009}]{2009A&A...496..577Z}
{Zechmeister} M.,  {K{\"u}rster} M.,  2009, \mn@doi [\aap] {10.1051/0004-6361:200811296}, \href {https://ui.adsabs.harvard.edu/abs/2009A&A...496..577Z} {496, 577}

\bibitem[\protect\citeauthoryear{{van Dokkum}}{{van Dokkum}}{2001}]{2001PASP..113.1420V}
{van Dokkum} P.~G.,  2001, \mn@doi [\pasp] {10.1086/323894}, \href {https://ui.adsabs.harvard.edu/abs/2001PASP..113.1420V} {113, 1420}

\makeatother
\end{thebibliography}



\appendix
\section{APPENDIX}

\begin{figure}
\centering
\hspace*{0.65cm}\includegraphics[trim={0 0.0cm 0.0cm 0},scale=0.83]{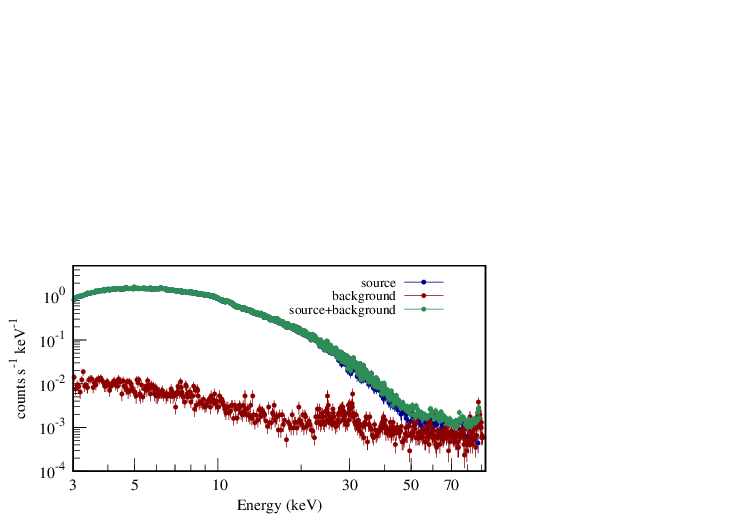}\vspace*{-3.5cm}
\hspace*{0.65cm}\includegraphics[trim={0 0.0cm 0.0cm 0},scale=0.83]{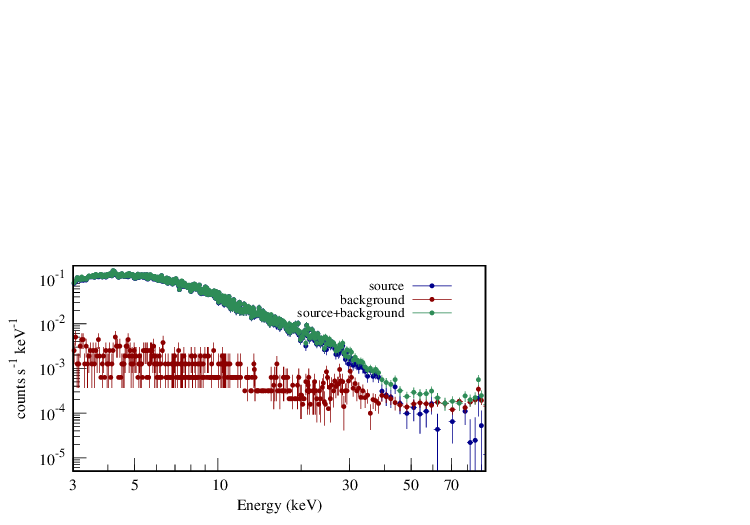}
\caption{  The top and bottom panels show a comparison of source (blue), background (brown), and source+background (green) count rates for NuSTAR ID~02 and ID~04, respectively}
\label{fig:id_04_source_back}
\end{figure}

\begin{table*}
\scriptsize
\caption{ Investigation of observed (source+background) and background count rates in different energy ranges for ID~02 and ID~04.}
 \label{tab:src_back_count}
\centering
\begin{tabular}{cccccc}
\hline
\hline
Instrument               &  40-50 keV     &  50-60 keV        &40-60 keV             & 60-70 keV        & 70-79 keV       \\ [4pt]
\hline 
ID 02\\ [4pt]
\hline
FPMA observed (10$^{-2}$)  &4.04$\pm$0.10  &2.00$\pm$0.07   & 6.13$\pm$0.12        &1.56$\pm$0.06     & 1.16$\pm$0.06   \\[6 pt]
FPMA bkg (10$^{-2}$)       &0.78$\pm$0.04  &0.76$\pm$0.04   & 1.54$\pm$0.06        &0.65$\pm$0.04     & 0.55$\pm$0.04   \\[6 pt]
FPMB observed (10$^{-2}$)  &4.00$\pm$0.10  &2.25$\pm$0.07   & 6.33$\pm$0.12        &1.68$\pm$0.06     & 1.04$\pm$0.05   \\[6 pt]
FPMB bkg (10$^{-2}$)       &0.07$\pm$0.04  &0.64$\pm$0.003  & 1.33$\pm$0.055       &0.61$\pm$0.04     & 0.50$\pm$0.03   \\[6 pt]
\hline 
ID 04 \\ [4pt]
\hline
FPMA observed (10$^{-3}$)  & 2.95$\pm$0.27  & 1.55$\pm$0.20     & 5.24$\pm$0.44        &1.52$\pm$0.20     & 0.70$\pm$0.10   \\[6 pt]
FPMA bkg (10$^{-3}$)       & 1.67$\pm$0.20  & 1.59$\pm$0.19     & 3.26$\pm$0.28        &1.57$\pm$0.19     & 1.35$\pm$0.18   \\[6 pt]
FPMB observed (10$^{-3}$)  & 2.97$\pm$0.30  & 1.49$\pm$0.20     & 5.21$\pm$0.36        &0.74$\pm$0.14     & 0.74$\pm$0.20   \\[6 pt]
FPMB bkg (10$^{-3}$)       & 1.66$\pm$0.20  & 1.66$\pm$0.20     & 3.32$\pm$0.28        &1.14$\pm$0.17     & 1.51$\pm$0.19   \\[6 pt]
\hline
\end{tabular}
\end{table*}

\begin{table*}
\scriptsize
\setlength{\tabcolsep}{5pt}
\caption{Broadband spectral fitting of ID 02 with different continuum models in 3-40 keV (without GABS) and 3-79 keV energy band (with GABS).}
 \label{tab:specfit002}
\centering
\begin{tabular}{l@{\hspace{0.5cm}}lcccccccc} 
\hline
Models     &Parameters                    &CUTOFFPL                 &CUTOFFPL+GABS           &HIGHECUT                &HIGHECUT+GABS          &FDCUT                   
           &FDCUT+GABS                    &NPEX                     &NPEX+GABS                  \\ [4pt]
\hline
TBabs      &$\rm N_{H} (10^{22}~cm^{-2}$) &$0.43^{+0.04}_{}$        &$0.43^{*}_{}$           &$2.37^{+0.18}_{-0.18}$  &$1.4^{+0.2}_{-0.2}$    &$1.20^{+0.20}_{-0.20}$ 
           &$1.34^{+0.20}_{-0.20}$        &$0.43^{+0.03}_{}$        &$0.43^{*}_{}$              \\  [6 pt]
Powerlaw   &Photon Index ($\Gamma$)       &$0.11^{+0.05}_{-0.05}$   &$0.91^{+0.07}_{-0.07}$  &$1.33^{+0.01}_{-0.01}$  &$1.24^{+0.02}_{-0.01}$ &$1.20^{+0.05}_{-0.05}$            
           &$1.23^{+0.01}_{-0.01}$        &$0.06^{+0.04}_{-0.04}$   &$0.74^{+0.03}_{-0.03}$     \\  [6 pt]
           &$\rm E_{cut}$ (keV)           &$10.72^{+0.30}_{-0.30}$  &$38.1^{+3.6}_{-4.9}$    &$18.6^{+0.3}_{-0.3}$    &$35.4^{+2.5}_{-1.4}$   &$29.10^{+0.50}_{-0.50}$          & $71.6^{+4.9.1}_{-7.1}$       &$10.3^{+0.2}_{-0.2}$     &$21.2^{+0.82}_{-1.1}$       \\  [6 pt]
           &$\rm E_{fold}$ (keV)          & --                      & --                     &$19.7^{+0.5}_{-0.5}$    & $36.3^{+6.5}_{-6.5}$  &$7.90^{+0.30}_{-0.30}$           &$12.8^{+6.6}_{-4.5}$          & --                      & --                        \\ [10pt]
           & Normalization (10$^{-2}$)    &$0.85^{+0.08}_{-0.07}$   & $3.13^{+0.29}_{-0.16}$ &$5.80^{+0.10}_{-0.10}$   & $4.90^{+0.15}_{+0.10}$ & $4.77^{+0.12}_{0.12}$
           & $4.91^{+0.14}_{0.12}$        &$0.80^{+0.06}_{-0.07}$   &  $2.51^{+0.13}_{-0.11}$    \\ [10pt]
           &f$_{\text{NPEX}}$ (10$^{-4}$) &--                       & --                     &--                      & --                    & --
           & --                           &$0.4^{*}$                &  $0.1^{+0.01}_{-0.03}$     \\ [10pt]
Blackbody  &Temp. (keV)                   &$1.06^{+0.02}_{-0.02}$   &$1.09^{+0.06}_{-0.05}$  & --                     & --                    & --                              
           & --                           &$1.07^{+0.02}_{-0.02}$   &$1.03^{+0.03}_{-0.03}$   \\ [6 pt]
           &Norm                          &$8.31^{+0.40}_{-0.40}$   &$2.11^{+0.32}_{-0.29}$  & --                     & --                    & --                              & --                           &$8.5^{+0.4}_{-0.4}$      &$3.9^{+0.2}_{-0.4}$        \\ [10pt] 
Gaussian  &E (keV)                        &$6.31^{+0.06}_{-0.06}$   &$6.29^{+0.06}_{-0.06}$  &$6.28^{+0.06}_{-0.06}$  &$6.28^{+0.07}_{-0.06}$ &$6.28^{+0.06}_{-0.06}$           
          &$6.28^{+0.7}_{-0.05}$          &$6.29^{+0.06}_{-0.07}$   &$6.29^{+0.06}_{-0.06}$     \\ [6 pt] 
          &Eqwidth (eV)                   &$20.2^{+5.1}_{-5.3}$     &$18^{+5}_{-4}$          &$17^{+4}_{-4}$          &$16^{+4}_{-4}$         &$16^{+5}_{-4}$                   &$16^{+5}_{-4}$                 &$17^{+5}_{-4}$           &$19^{+5}_{-4}$             \\ [10pt]
GABS      &E$_{CRSF}$ (keV)               & --                      &$50.4^{+0.2}_{-0.4}$    & --                     &$48.7^{+1.0}_{-1.2}$   & --                              
          &$50.8^{+0.4}_{-0.4}$           & --                      &$50.7^{+0.6}_{-0.5}$       \\ [6 pt] 
          &$\sigma_{CRSF}$ (keV)          & --                      &$13.6^{+1.3}_{-1.3}$    & --                     &$13.9^{+0.6}_{-0.6}$   & --                              &$14.2^{+0.4}_{-0.4}$           & --                      &$12.6^{+0.7}_{-1.1}$       \\ [6 pt]
          &Strength$_{CRSF}$ (keV)        & --                      &$50.5^{+13.4}_{-7.9}$   & --                     &$58.5^{+4.8}_{-5.2}$   & --                              
          &$62.9^{+2.3}_{-1.9}$           & --                      &$43.9.0^{+5.6}_{-6.5}$      \\ [10pt]
Luminosity (1-70 keV) &($\rm 10^{36}~erg~s^{-1}$) & --              &5.57                    & --                     &5.68                   & --                             
          &5.68                           & --                      &5.56                      \\ [6 pt] 
Fit Statistics &$\rm\chi_{red}^{2}/d.o.f.$&1.13/1652                &1.04/1890               &1.21/1653               &1.03/1890              &1.04/1653                        
          &1.04/1890                      &1.14/1652                &1.03/1890                 \\ [6 pt]
\hline \\
\end{tabular}\\
\hspace{-13cm}(*)= parameter fixed while calculating error\\
\hspace{-12.3cm}Note= Galactic N$_{\text{H}}$ of source is $\approx$0.43$\times$10$^{22}$ cm$^{-2}$\\
\hspace{-10cm}f$_{\text{NPEX}}$ = fraction parameter (A$_{\text{p}}$/A$_{\text{n}}$; see Equation 6 of \citealt{1999ApJ...525..978M})

\end{table*}

\begin{table*}
\scriptsize
\caption{Best-fit parameters obtained from the spectral analysis of NuSTAR \& NICER observations of IGR~J06074+2205 on 60232 MJD for observation id 90901331004 using NPEX as the continuum model. The distance of the source used for luminosity calculation is 5.99 kpc.}
 \label{tab:specfit_nicer_nustar}
\centering
\begin{tabular}{ ccc } 
\hline
MJD               &                                            & 60232                      \\
\hline
Model             & Parameters                                 & NICER+NuSTAR               \\ [4 pt]
Energy Range      &                                            & 1-50 keV                   \\ [4 pt]
\hline 
TBabs             & $\rm N_{H} (10^{22}~cm^{-2}$)              & $1.25_{-0.08}^{+0.09}$     \\ [6 pt]
NPEX              & Photon Index ($\Gamma$)                    & $1.36_{-0.03}^{+0.03}$     \\ [6 pt]
                  & $\rm E_{cut}$ (keV)                        & $7.64_{-0.30}^{+0.30}$     \\ [6 pt]
                  & Normalization (10$^{-2}$)                  & $1.14_{-0.06}^{+0.06}$     \\ [6 pt]
                  & f$_{\text{NPEX}}$ (10$^{-4}$)              & $1.30_{-0.20}^{+0.20}$     \\ [6 pt]
Luminosity (1-70 keV) & ($\rm 10^{36}~erg~s^{-1}$)             & 0.47                       \\ [4 pt] 
Fit Statistics    & $\rm \chi_{red}^{2}/d.o.f.$                & 1.03/933                   \\
\hline
\end{tabular}
\end{table*}
\bsp	
\label{lastpage}
\end{document}